\documentclass[preprint]{elsarticle}
\usepackage{graphicx}
\usepackage{dcolumn}
\usepackage{bm}
\usepackage[T1]{fontenc}
\usepackage[cp852]{inputenc}
\usepackage[a4paper]{geometry}
\usepackage{float}
\usepackage{units}
\usepackage{textcomp}
\usepackage{amsmath}
\usepackage{amssymb}
\usepackage{esint}
\usepackage{amsmath}   
\usepackage[all]{xy}
\usepackage{multirow, array}
 \usepackage[titletoc]{appendix}
\usepackage{rotating}

\DeclareGraphicsExtensions{.jpg}

\begin{document}

\title{A second binding model to study diffusion of $Cr$ diluted in BCC $Fe$.}
\begin{frontmatter}

\author{Viviana P. Ramunni}
\ead{vpram@cnea.gov.ar}
\address {CONICET - Avda. Rivadavia 1917, Cdad. de Buenos Aires, C.P. 1033, Argentina.}  
\address {Departamento de Materiales, CAC-CNEA, Avda. General Paz 1499, 1650 San Mart{\'i}n, Argentina. \\
 Phone number: +54 11-6772-7298; Fax: +54 11-6772-7303}
\thanks{This work was partially financed by CONICET - PIP 00965/2010.}
\date{\today}

\begin{abstract} 
A classical molecular static technique (CMST) and DFT calculations using SIESTA, are employed here to characterize the self diffusion and the tracer solute diffusion in the bulk of BCC diluted $FeCr$ alloy driven by both vacancy and interstitial migration. For the first time in the literature, a six-frequency model (developed by Okamura and Allnatt) involved in a second nearest neighbor binding approach is adapted for calculations in a real system. We obtain microscopic parameters, namely: i) the free energy of vacancy formation and the vacancy-solute binding energy, ii) the involved jump frequencies, and  iii) the tracer correlation factor.  The present approximation describes much better the experimental data of self and solute atoms than recent calculations using a first binding approach. Also, we confirm that a vacancy drag mechanism is unlikely to occur in $FeCr$ diluted alloys. Our results also show that the diffusion processes is mainly mediated by vacancies, while diffusion by intertitial mechanism is several orders of magnitudes slower. 
\end{abstract}


\end{frontmatter}

\section{Introduction}

Recently, Hurtado \textit{et al} \cite{HUR10,HUR11} have studied the hydrogen ($H$) diffusion effects on $9Cr$ steels, presently used in conventional supercritical thermal power plants and candidates for future IV generation supercritical water cooled nuclear reactors. Based on the numerical resolution of Fick's equations in presence of trapping sites \cite{CAS12}, and from the fit of electrochemical $H$ detection curves, these authors provided quantitative information about the binding energy between $H$ and trapping sites during $H$ diffusion process \cite{HUR10}. Permeation tests performed in our laboratory on $9Cr$ alloys reveal a permeation coefficient 10 times lower and a diffusion coefficient 200 times lower than in pure, annealed iron. Focusing on these experimental results, we will explore very simple model of new $H$ trapping sites and possible migration paths that can explain the experimental observations. 

Therefore, before dealing with modeling $H$ traps, it is important to watch carefully and with special attention the initial microscopic processes that can delay the $H$ atoms during diffusion. In this way, we start studying numerically the static and dynamic properties of vacancies and interstitial defects in a $FeCr$ matrix. $9Cr$ Martensitic steel of BCT structure, is a complex system to be simulated. Our main difficulty arises in modeling such a system. For this purpose, we have considered an incremental approach, namely: the effect of the ratio between the BCT cell parameters ($c/a$) is not as relevant as the effect of substitutional $Cr$ \cite{HUR10}, then we start our calculation in $\alpha-Fe$ bulk in presence of: i) a vacancy at nearest neighbor sites of a substitutional $Cr$ atom; followed by the incorporation of ii) a single interstitial $Cr$. Both defective systems are studied separately in the present work.

We use a classical molecular statics technique (CMST) and DFT calculations using the SIESTA code \cite{ART99} coupled to the Monomer method \cite{RAM06,RAM09,RAM09b,RAM11}. the Monomer method coupled to CMST simulations is a much less computationally expensive method, that allows us to compute at low cost a bunch of jump frequencies from which we can perform averages in order to obtain more accurate effective frequencies. On the other hand, when the Monomer is coupled to DFT calculations, the method is akin to the Dimer one from the literature \cite{HEN01}, but roughly employs half the number of force evaluations which is a great advantage in SIESTA ab-initio calculations. 

We proceed as follows: we calculate the full set of frequencies employing the economic Monomer method \cite{RAM06}. The Monomer \cite{RAM06} is used to compute the saddle points configurations from which we obtain the jumps frequencies defined in the six-frequency model, in the context of a second nearest neighbors binding model developed by Okamura and Allnatt \cite{OKA83}. Then, the full set of phenomenological coefficients are obtained in terms of these six-frequencies. In the frame of the non-equilibrium thermodynamics, we obtain the flux equations from which we express analytic expressions for diffusion coefficients. 

Recently, similar approaches than our have been been performed.
For BCC structures, although using a first nearest neighbor shell approximation of the binding between vacancies and solute atoms, was performed by Choudhury \textit{et al.} \cite{CHO11} using VASP. The authors have calculated the self-diffusion and solute diffusion coefficients in diluted $\alpha FeNi$ and $\alpha FeCr$ alloys including an extensive analysis of the phenomenological Onsager $L$-coefficients for describing segregation.

Also, in a first shell binding model, but for FCC structures, we have presented  studies of impurity diffusion behavior in Nickel-Aluminium and Aluminium-Uranium diluted alloys \cite{RAM13,RAM14}. There we show how the CMST is appropriate in order to describe the impurity diffusion behavior mediated by a vacancy mechanism. In Ref. \cite{RAM14}, our results with CMST show excellent agreement with those from molecular dynamic calculations for the $U$ mobility diluted in $Al$. While, in \cite{RAM13}, CMST calculations performed  in diluted $NiAl$ and $AlU$ f.c.c. alloys shows excellent agreement  with available experimental data for both systems. Also, in \cite{RAM13} we have reported for the first time the behavior of diffusion coefficients for the solute-vacancy paired specie,  predicting that for $NiAl$ the solute diffuses through a vacancy interchange mechanism, while for the $AlU$ system, a vacancy drag mechanism occurs.

In the present work, for CMST calculations the inter-atomic interactions are represented by suitable EAM potentials \cite{RAS09}, while for DFT calculations, we employ the generalized gradient approximation (GGA) for exchange and correlation for $Fe$ and $Cr$. For diffusion process mediated by vacancies we employ the six-frequency model, in the context of a second nearest neighbors binding model developed by Okamura and Allnatt \cite{OKA83}. It must be noted that this is the first time that such a powerful procedure is employed for real systems. Our CMST calculations for vacancies, reveal that $Cr$ in $Fe$ at diluted concentrations migrates as free species, result that is confirmed by our DFT calculation using SIESTA. In this respect we conclude that a vacancy drag mechanism is unlikely to occur in BCC $FeCr$ in accordance with \cite{CHO11}. 

For the case of diffusion process mediated by interstitials in BCC $FeCr$ matrix, we use the full set of Onsager $L$-coefficient developed by Barbe and Nastar \cite{BAR07}. Our results also show that the diffusion coefficients for the intertitial process are several orders of magnitudes lower than those mediated by vacancies. This last ones are in good agreement with experimental available data. Then,  we can conclude  that the diffusion processes is mainly mediated by vacancies.

The paper is organized as follows. In Section \ref{S1}, for diffusion mediated by a vacancy mechanism, we briefly introduce the full set of the Onsager coefficients calculated by Okamura and Allnatt \cite{OKA83} in the second-nearest neighbor binding model for BCC lattices. Analytic expressions of the diffusion coefficients in binary alloys and the corresponding one for the paired species, $Cr+V$, all of them in terms of Onsager coefficients, are presented in section \ref{S2}. This allows to express the diffusion coefficients in terms of the frequency jumps. Section \ref{S3}, is devoted to present our numerical results for vacancies using the theoretical procedure previously summarized and making a comparison with available experimental data. For interstitials, we will not not show here, but we use the theoretical procedure developed by Barbe and Nastar \cite{BAR07}. The last section briefly presents some conclusions.

\section{A second binding model for diffusion mediated by vacancies in \textbf{BCC} lattices}
\label{S1}

In the framework of linear non-equilibrium thermodynamics, Okamura and Allnat \cite{OKA83} using symmetry types by a suitable classification of the vacancy-A-atom exchange, have obtained analytical expressions for the phenomenological $L-$ coefficients, from which transport phenomena can be described. In particular, atomic diffusion in $FeCr$-alloys is described through the following parameters, namely: $L_{FeFe}$, $L_{FeCr}=L_{CrFe}$ and $L_{CrCr}$ phenomenological coefficients, $D^{\star}_{Fe}$ and $D^{\star}_{Cr}$ the self and tracer solute diffusion coefficients for the solvent $Fe$ and solute $Cr$ species. In this context, the flux $J_i$ of the component $i$ relative to the local crystal lattice in an isothermal isotropic crystal is given by
\begin{equation}
J_i=-\sum_j L_{ij}(\nabla \mu_j)_T,
\label{ji0}
\end{equation}
when there are not forces acting on the system. In (\ref{ji0}) $\mu_j$ is the chemical potential of component $j$ and $L_{ij}$ are the phenomenological coefficients. The main idea of present calculations is to focus attention on transport phenomena of diluted binary alloys, specifically on the corresponding solute transport coefficient and how this is related to the diffusion coefficient through the flux equations (\ref{ji0}). In this way, Okamura and Allnatt give compact expressions for the $L-$coefficients by treating the contribution of the vacancy-impurity exchanges as a perturbation in evaluating the Green function by inversion of a certain reduced matrix $Q$ (which will be defined below). 

In the following, we present the Okamura and Allnat expressions for the $L-$coefficients in the particular case of BCC lattices \cite{OKA83}, in terms of the frequency rates in the context of the six frequency model \cite{LEC78}. In this way, the effect of different vacancy exchange mechanisms on solute diffusion can easily be understood. We assume that the perturbation of the solute movement by a vacancy $V$, is limited to its immediate vicinity. Established the $Mth$-nearest-neighbor binding model, we identify the frequencies as follows \cite{OKA83}:
\begin{itemize}
\item (i) $\omega _2$ for exchange with a $Cr$ atom,
\item (ii) $k_{nm}$ for exchange with a $Fe$ atom when initially the $Fe$ is an $nth$-n.n. and the vacancy is an $mth$-n.n. to a $Cr$ atom ($n,m \leq M$).
\item (iii) $k_{0m}$ for exchange with a $Fe$ that is more distant than $Mth$-n.n. when the vacancy is an $mth$-n.n. neighbor to a $Cr$ ($m\leq M$).
\item (iv) $k_{n0}$ for exchange with a $Fe$ that is an $nth$-n.n ($n\leq M$) when the vacancy is more distant than the $Mth$-n.n. neighbor to a $Cr$.
\item (v) $\omega _0$ for exchange with a $Fe$ when both the atom and the vacancy are more distant than $Mth$-n.n. to a $Cr$ atom.
\end{itemize} 
\begin{figure}[h]
\begin{center}
\includegraphics[width=10.0cm]{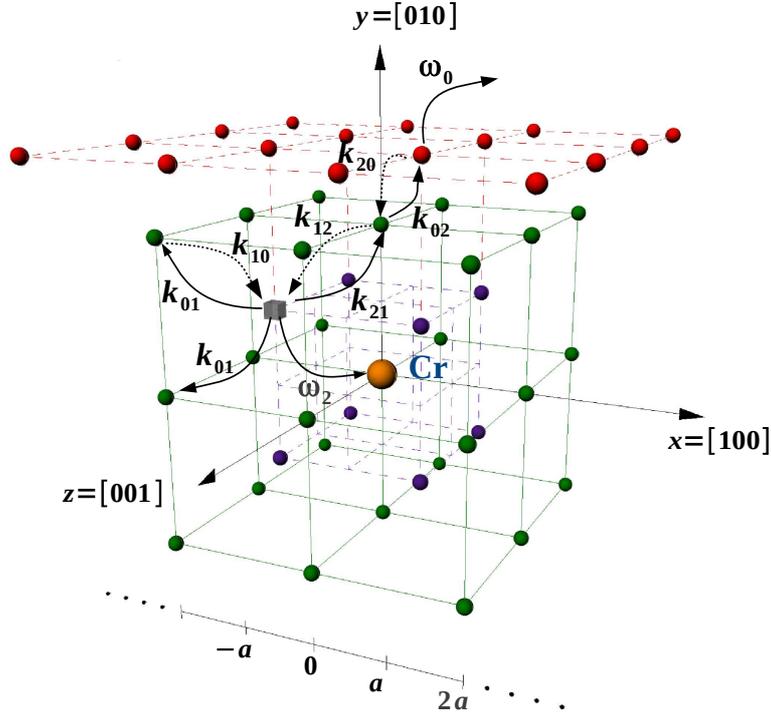}
\caption{The six-frequencies involved in the second binding model for BCC lattices. In figure, blue bullets are the first eight neighbors sites to the solute $\bf{S}$ at the origin. In green the 26 subsequent sites. In red, the third coordinated shell from which the vacancy never returns to the second shell.}
\label{6f-V}
\end{center}
\end{figure}
If $M=2$ then, six frequencies are assigned in the Type I Le Claire Model \cite{JON72}, namely:
($\omega_2$,$k_{12}$,$k_{21}$,$k_{01}$,$k_{02}$,$\omega_0$). Detailed balance requires the relation
\begin{equation}
\frac{k_{01}}{\omega_0} = \exp(-\beta \xi_1) \, \, \, ; \, \, \, 
\frac{k_{02}}{\omega_0} = \exp(-\beta \xi_2) \, \, \, ; \, \, \, 
\frac{k_{21}}{k_{12}} = \frac{k_{01}}{k_{02}} = \exp(\beta (\xi_2-\xi_1))  
\end{equation}

The reduced matrix $Q$ can be written as,
\begin{eqnarray}
Q & = & \left (\begin{array}{ccc}
-(\omega_2 + 3k_{21}+4k_{01}) & k_{12} & W1 \\
4k_{21} & -(4k_{12} + 4k_{02}) & W2\\
2k_{01} & 0 & W3 \\
0 & k_{02} & W4 \\
0 & 0 & W5 \\
k_{01} & 0 & W6 \\
0 & 0 & W7  
\end{array} \right )
\label{Q}
\end{eqnarray}
where $W_i$, has elements that are multiples of $\omega_0$ \cite{OKA83},
\begin{eqnarray}
W_1 &=& (2\omega_0,0,0,\omega_0,0,...) \\  
W_2 &=& (0,4\omega_0,0,0,0,...) \\
\vdots
\end{eqnarray}

The zero-frequency limit of the linear response formula for the phenomenological Onsager coefficients $L_{ij}$ taken from Okamura and Allnatt\cite{OKA83} work, implies
\begin{equation}
L_{CrCr}=A\omega_2f_{Cr},
\label{LCrCr}
\end{equation}
\begin{equation}
A=\beta Na_{Fe}^2C_{p(1)}n_1/z_1, 
\label{ACr}
\end{equation}
where $n_1$ is the number of sites in symmetry type $1$ ($n_1=n_{\overline{1}}$), $N$ is the number of sites per unit volume, and the impurity correlation factor, $f_{Cr}$, is given by
\begin{equation}
f_{Cr}=\frac{7k_{01}F}{(2\omega_2+7k_{01}F)}.
\label{fCr}
\end{equation}

Introducing (\ref{ACr}) and (\ref{fCr}) in (\ref{LCrCr}), the final expression for $L_{CrCr}$ coefficient is,
\begin{equation}
L_{CrCr}= \frac{\beta Na_{Fe}^2C_{p(1)}n_1}{z_1} \times \frac{7k_{01}F}{(2\omega_2+7k_{01}F)}.
\label{LCrCrf}
\end{equation}

As in Ref. \cite{RAM13}, it is essential to obtain the reduced matrix $Q$ from the kinetic equations in term of the jump frequencies, then the factor $F$ introduced by Manning \cite{MAN68} is calculated from $Q$ in (\ref{Q}) as,  
\begin{equation}
-Q_{11}/|Q|=(\omega_2+7k_{01}F)^{-1} \Rightarrow 7F=(2x^2+B_1x+B_2)/(x+B_3),
\label{Q11}
\end{equation}
where $x=k_{21}/k_{01} = k_{12}/k_{02}$, the constants $B_n$ are numbers that can be calculated numerically and values are summarized in Table \ref{Bn}.

For $L_{FeCr}=L_{CrFe}$ and $L_{FeFe}$, we have
\begin{equation}
L_{FeCr} = -2A\frac{\omega_2 k_{01}F_1}{2\omega_2+7k_{01}F} \,\,\,  ,
\end{equation}
where
\begin{equation}
F_1=(x-2)+(x-1)\left(\frac{x+B_4}{x+B_3}\right).
\label{F1}
\end{equation}
and $L_{FeFe}$ is decoupled in two parts as,
\begin{equation}
L_{FeFe}=L^{(0)}_{FeFe}+L^{(1)}_{FeFe},
\end{equation}
\begin{equation}
L^{(0)}_{FeFe}=A(6k_{21}+7k_{01}) + 4\beta a_{Fe}^2N\omega_0C^{\prime}_V \,\,\, ,
\end{equation}
and completing the full set of $L$-coefficients,
\begin{equation}
L^{(1)}_{FeFe}= -\frac{2A}{(2\omega_2+7k_{01}F)}\left\{k_{01}(k_{01}F_2+2\omega_2F_3) + \frac{\omega_2k_{01}^2F_4}{\omega_2+7k_{01}F} \right\}.
\end{equation}
Expressions for $F_i$ from Ref. \cite{OKA83} are,
\begin{eqnarray}
F_2&=&(x-2)^2+[(x-2)(x-1)(2x+B_4+B_5)/(x+B_3)]+(x-1)^2(3x+B_6)/(x+B_3) \nonumber \\  
F_3&=&(x-1)^2/(x+B_3)  \label{Fj}\\  
F_4&=&B_7(x-1)^2/(x+B_3)^2.  \nonumber  
\end{eqnarray}
The coefficients $B_6$ and $B_7$ in $F_i$ are calculated as,
\begin{eqnarray}
B_6&=& B_1-3B_3+B_4+B_5\\  
B_7&=& B_3B_6-B_4B_5-B2.  \nonumber  
\end{eqnarray}
Coefficients $B_i$ ($i=1-7$) in equations (\ref{Q11}), (\ref{F1}) and (\ref{Fj}) are present in Table \ref{Bn} below,
\begin{table}[h]
\begin{center}
\caption{Numerical constants $B_n$ for the second-nearest-neighbor binding model taken from Refs. \cite{OKA83,MAN68}.}
\label{Bn}
\begin{tabular}{c|ccccccc}
\hline 
\, \, & \, $B_1$ \, & \, $B_2$ \, & \, $B_3$ \, & \, $B_4$ \, & \, $B_5$ \, & \, $B_6$ \, & \, $B_7$ \,
\tabularnewline
\hline 
\hline 
\,  Ref.\cite{OKA83} \, & \, 5.175 \, & \, 2.466 \, & \, 0.8082 \, & \, 0.1713 \, & \, 0.1713 \, & \, 3.093 \, & \, 0.0044 \,
\tabularnewline 
\hline 
\,  Ref.\cite{MAN68} \, & \, 5.182 \, & \, 2.476 \, & \, 0.8106 \, & \, - \, & \, - \, & \, - & \, - \,
\tabularnewline 
\hline 
\end{tabular}
\end{center}
\end{table}

We shall need the concentration, $c_{P(m)}$, of $mth$-nearest neighbors (n.n.) pairs ($m\leq M$) given by the law of mass action
\begin{equation}
C_{p(m)}/C^{\prime}_{Cr}C^{\prime}_V=z_m \exp(\beta \xi_m),
\label{CPM}
\end{equation}
where $z_m$ is the number of $mth$-n.n. sites of any site ($z_m=8$ for \textbf{BCC} structures), $C^{\prime}_{Cr}$, $C^{\prime}_V$ are respectively the concentration of free impurities and vacancies, e.g.
\begin{equation}
C^{\prime}_{Cr}=C_{Cr} - \sum_{m=1}^M C_{p(m)},
\end{equation}
and
\begin{equation}
C^{\prime}_{V}=C_{V} - \sum_{m=1}^M C_{p(m)},
\end{equation}
where $\beta = 1/k_BT$ the absolute temperature, $\epsilon _m$ is the energy of interaction of a vacancy and an impurity at $mth$-n.n. separation and $\xi _m$ is the binding energy for a $Mth$-n.n. binding model defined as,
\begin{eqnarray}
\xi_m  & = & \left\{\begin{array}{ccc}
 - \epsilon_m & ; &  m=1,2,...,M \\
  0 & ; & otherwise.
\end{array} \right.
\label{Ebind}
\end{eqnarray}
In the case of $M=2$, which corresponds to the second-nearest-neighbor binding model, we shall need analytical expression for $C_p(1)$, $C_p(2)$ and $C^{\prime}_V$ and $C^{\prime}_{Cr}$ in terms of the lattice parameters, the vacancy formation energy in perfect lattice and the binding vacancy-solute energies at $m=1,2$ separations. In metals $C^{\prime}_{Cr} \simeq C_{Cr}$ and we can express $C_p(m)$ in terms of the known molar concentrations $C_{Cr}$ and $C_V$ as follows, 
\begin{eqnarray}
C_{p(1)} &=&C^{\prime}_{V}C^{\prime}_{Cr}z_1 \exp (\beta\xi_1), \\
C_{p(2)} &=&C^{\prime}_{V}C^{\prime}_{Cr}z_2 \exp (\beta\xi_2)
\label{cp12}
\end{eqnarray}
and 
\begin{equation}
C^{\prime}_{V}=C_V-\left\{C_{p(1)}+C_{p(2)}\right\}.
\label{cvp}
\end{equation}
Defining the equilibrium constants $K_{(1)}$, $K_{(2)}$ as
\begin{eqnarray}
K_{(1)} &=&z_1C_{Cr}\exp(\beta\xi_1), \label{k12}\\
K_{(2)} &=&z_2C_{Cr}\exp(\beta\xi_2)
\label{cp13}
\end{eqnarray}
and introducing (\ref{cp12}) and (\ref{cp13}) in (\ref{cvp}) we obtain,
\begin{equation}
C^{\prime}_{V}=\frac{C_V}{\left\{ K_{(1)}+K_{(2)} \right\}}.
\label{cvp2}
\end{equation}
and
\begin{equation}
C_{p(1)}=\frac{C_VK_{(1)}}{1+\left\{ K_{(1)}+K_{(2)} \right\} }.
\label{cp2}
\end{equation}

In the context of the Transition State Theory, we assume that the jumps are thermally activated and then $k_{ij}$ can be expressed as,
\begin{equation}
k_{ij}=k_0(T)\exp(-E^{\rightarrow}_m/k_BT).
\label{nu0T}
\end{equation}
$k_{00}=\omega _0$, $k_{22}=\omega _2$ and $k_{ij}$ depends on the migration barriers $E^{\rightarrow}_m$ which are calculated using the Monomer Method \cite{RAM06}. The migration barriers $E^{\rightarrow}_m$ are summarized in Table (\ref{T5}). We adopt a temperature dependent attempt frequency $k_0(T)$ in terms of the Wert model\cite{WER65} can be written as
\begin{equation}
k_0(T)=\frac{k_B T}{h}.
\label{nuT}
\end{equation}
We also assume a constant attempt frequency of $k_0=5\times 10^{12}Hz$ taken from Ref. \cite{CHO11}.  

\section{Expressions for $D^{\star}_{Fe}$, $D^{\star}_{Cr}$ and $D_{(Cr+V)}$ coefficients}
\label{S2}
A comparison between experimental data and present simulations are possible with the knowledge of the two tracer diffusion coefficients $D^{\star}_{Fe}$ and $D^{\star}_{Cr}$. The tracer self-diffusion coefficient $D^{\star}_{Fe}$ of $Fe$ in a diluted alloy with a concentration $C_{Cr}$ of solute atoms, can be written in terms of the self diffusion coefficient $D^{\star}_{Fe}(0)$, as
\begin{equation}
D^{\star}_{Fe}(C_{Cr})=D^{\star}_{Fe}(0)(1+b_{{Fe}^{\star}}C_{Cr}),
\label{DAenh}
\end{equation}
at first order in $C_{Cr}$. The solvent enhancement factor, $b_{Fe}$, is obtained in terms of the properties of the solute-vacancy model. On the other hand, for the pure solution, the self diffusion coefficient $D^{\star}_{Fe}(0)$ is given by \cite{LEC78},
\begin{equation}
D^{\star}_{Fe}(0)=a_{Fe}^2c^{0}_Vf_0\omega_0.
\end{equation}
where $a_{Fe}$ is the solvent lattice parameter, $f_0=0.7272$ is the correlation factor for the self-diffusion in BCC lattices, and $c^0_V$ is the vacancy concentration at the thermodynamical equilibrium. This former is such that,
\begin{equation}
c^{0}_V=\exp\left( -\beta E^V_f\right),
\label{cv0}
\end{equation}
where $\beta=1/k_BT $ is the absolute temperature, $E^V_f$ is the formation energy of the vacancy in pure $Fe$. The entropy terms are set to zero, which is a simplifying approximation.
So that, inserting (\ref{cv0}) we get 
\begin{equation}
D^{\star}_{Fe}(0)=a_{Fe}^2f_0\omega_0\exp\left( -\beta E^V_f\right).
\label{DFe0}
\end{equation}
\noindent We assume $C_{Cr} \rightarrow 0$ then, we use pure lattice parameters for all our calculations. The solute-enhancement factor $b_{Fe^{\star}}$, is taken from Ref. \cite{LEC78} for the Type I-BCC model is expressed as,
\begin{eqnarray}
b_{Fe^{\star}} & = & -20 +14\left(\frac{\mu_1}{f_0}\right)+6\left(\frac{k_{21}}{\omega_0}\right)\left( \frac{\nu _1}{f_0}\right) \exp(-\beta \xi_1),
\label{bFe}
\end{eqnarray}
where $\mu_1$ and $\nu_1(\omega_2/k_{12},k_{21}/k_{12})$ are mean partial correlation factors and known functions of the frequency ratios in the parentheses \cite{JON72}. 


In the particular case of a binary dilute $FeCr$ alloy containing $N_{Fe}$ solvent atoms, $N_{Cr}$ solute atoms and $N_{V}$ vacancies, the flux of $Cr$ atoms is equal to
\begin{equation}
J_{Cr}=-\left(\frac{L_{CrCr}}{C_{Cr}}-\frac{L_{FeCr}}{C_{Fe}}\right)k_{B}T\left(1+\frac{\partial ln\gamma_{Cr}}{\partial \ln C_{Cr}}\right)\nabla C_{Cr}.
\label{JCr}
\end{equation}
where the coefficient $\gamma _{Cr}$ is defined in terms of the $Cr$ activity $a_{Cr}$. In the spite of the first Fick's law and assuming $\frac{\partial ln\gamma_{Cr}}{\partial \ln C_{Cr}}=0$, the expression for the tracer solute diffusion coefficient in the alloy is equal to that of the self solute diffusion coefficient (see Allnat and Lidiard \cite{ALL03}),
\begin{equation}
D_{Cr} = D^{\star}_{Cr} = \frac{k_{B}T}{N}\left(\frac{L_{CrCr}}{C_{Cr}}\right) \,\,\,\, ; \,\,\,\, C_{Cr} \rightarrow 0 \,.
\label{DCr}
\end{equation}

Introducing $L_{CrCr}$ from (\ref{LCrCrf}) and the detailed balance equation (\ref{CPM}) in (\ref{DCr}), we obtain an expression for $D^{\star}_{Cr}$ as,
\begin{equation} 
D^{\star}_{Cr}=a^2_{Fe}\omega_2\left(\frac{C_p}{3C_{Cr}}\right)\times \left\{\frac{7k_{01}F}{2\omega_2+7k_{01}F}\right\} \, = \, a^2_{Fe}\omega_2\left(\frac{C_p}{3C_{Cr}}\right)\times f_{Cr}.
\label{DBB2}
\end{equation}
The solute correlation factor $f_{Cr}$ from (\ref{DBB2}) is, 
\begin{equation}
f_{Cr}= \left\{\frac{7k_{01}F}{2\omega_2+7k_{01}F }\right\}.
\label{eq:FF1}
\end{equation}
where $F$ was previously defined in (\ref{Q11}). In the Le Claire description, $D^{\star}_{Cr}$ can also be expressed as,
\begin{equation}
D^{\star}_{Cr}=a_{Fe}^2f_{Cr}\omega_2 \exp \left[ -\beta(E^V_f+E_b)\right]
\label{DS22}
\end{equation}
For the drift of solutes in a vacancy flux we shall make contact with the alternative phenomenology offered by Johnson and Lam \cite{JOH76}. In terms of thermodynamic forces, which are precisely of the form required by non-equilibrium thermodynamics, up to second-virial coefficients, the flux of solute atoms $J_{Cr}$ is expressed as
\begin{equation}
J_{Cr}=-\sum _m D_{p(m)}\nabla C_{p(m)} + \sigma_V C^{\prime}_{Cr} D_V \nabla C^{\prime}_V,
\label{JP}
\end{equation}
The coefficients $D_p$ and $D_V$ are interpreted as diffusion coefficients of pairs and free vacancies, respectively, while $\sigma_V$ is a sort of cross section for vacancies to induce solute motion. We see that (\ref{JCr}) is equivalent to (\ref{JP}) if
\begin{equation}
D_{p(m)}=\frac{k_BT}{NC_{p(m)}}L_{CrCr}
\label{Dp}
\end{equation} 
For a vacancy mechanism, solute atoms may only move when they are paired with a vacancy and it is reasonable therefore that $D_{Cr}$ should be equal to ($C_p/C_{Cr}$) as (\ref{DCr}) and (\ref{Dp}) require. We proceed to show the results obtained by direct application of the previous theory, to the study of the diffusion of impurities in dilute alloys mediated by a vacancy mechanism. In the case of interstitials and in the second shell approximation, we employ the model developed by Barbe and Nastar \cite{BAR07} as in Ref. \cite{CHO11}. Here, we present a review of solute and solvent diffusion mediated by vacancies because it is a different approach that those used in recent literature \cite{CHO11,MAN09,RAM13}.


\section{Results}
\label{S3}
We present our numerical results for diluted $FeCr$ alloy. The interatomic interactions are represented by suitable $Fe$ and $Cr$ potentials \cite{RAS09}. We obtain the equilibrium positions of the atoms by relaxing the structure via the conjugate gradients technique. The lattice parameters that minimize the crystal structure energy are respectively $a_{Fe}=2.885\, $\AA \, and $a_{Fe}=2.865\, $\AA \, from DFT and classical methods. DFT calculations reported here were done with the ab-initio SIESTA \cite{ART99} code. We employ the generalized gradient approximation (GGA) for exchange and correlation for $Fe$ and $Cr$, a Mesh-CutOff parameter of 460Ry, a smearing temperature of 0.15eV (within a Fermi-Dirac scheme), and Brillouin zone sampling using a $7\times 7\times 7$k-points mesh for the bulk (which corresponds to a density of k-point of 26\AA /eV). A structure is considered relaxed when the forces are below $0.02eV$/ \AA. With this setup, we obtain a $0K$ lattice parameter for $\alpha Fe$ of 2.885 \AA. We use a default DZP basis for $Cr$ and another optimized one for $Fe$. All the calculations were performed at constant volume. We perform CMST and DFT calculations using respectively crystallite sizes containing $N=128,1024$ and $N=54,128$ of $Fe$ atoms. Eventually we include: i) one substitutional $Cr$ atom and a single vacancy, to study diffusion of $Cr$ dragged by vacancies and, ii) one interstitial $Cr$, which presents different relaxed configurations namely: mixed dumbbell, $Cr$ substitutional and a pure $Fe$ dumb-bell nearest neighbor to the $Cr$ which are summarized in Figure \ref{IntJ}. The current calculations have been performed at $T=0K$ and constant volume. In present calculations, the entropic barrier have been ignored.  
\subsection{Diffusion driven by vacancy migration}
\vspace{.5cm}
In Table \ref{T4} we establish a comparison between present calculation of formation and migration energies in perfecta BCC $Fe$ lattice from CMST and DFT (SIESTA+Monomer) calculations. As is usual, the vacancy formation energy ($E^V_f$) in pure $Fe$ is calculated as,
\begin{equation}
 E^V_f=E(N-1)+E_c-E(N),
\end{equation}
where, $E(N)$ for the perfect lattice of $N$ atoms, $E(N-1)$ is the energy of the defective system, and $E_c$ the cohesion energy. The migration barrier of the vacancy in perfect lattice ($E^V_m$), is calculated with the Monomer method \cite{RAM06}. For diluted alloys, we may consider the presence of the solute-vacancy complex $C_n=S+V_n$ in which, $n=1^{st}, 2^{nd}, 3^{rd},\dots$ (see the insets in Fig. \ref{T5}) indicates that the vacancy is a $n-$nearest neighbors of the solute atom
$S$. The binding energy between the solute and the vacancy for the complex $C_n=S+V_n$, defined in (\ref{Ebind}), is obtained as,
\begin{equation}
\epsilon^V_m=\left\{ E(N-2,C_n=Cr+V_n) + E(N)\right\} - \left\{ E(N-1,V) + E(N-1,Cr)\right\} ,
\label{Ebnn}
\end{equation}
where $E(N-1,V)$ and $E(N-1,Cr)$ are the energies of a crystallite containing ($N-1$) atoms of solvent $Fe$ plus one vacancy $V$, and one solute atom $Cr$ respectively, while $E(N-2,C_n=Cr+V_n)$ is the energy of the crystallite containing ($N-2$) atoms of $Fe$ plus one solute vacancy complex $C_n=Cr+V_n$. With the sign convention used here $\epsilon ^V_m < 0$ means an attractive solute-vacancy interaction, and $\epsilon ^V_m > 0$ indicates repulsion. We calculate the migration energies $E^{\rightarrow}_m,E^{\leftarrow}_m$ using the Monomer Method \cite{RAM06}, a static technique to search the potential energy surface for saddle configurations. The Monomer computes the least local curvature of the potential energy surface using only forces.

\begin{table}[ht] 
\begin{center}
\caption{Energies and lattice parameters for the pure BCC $Fe$ lattice. The first column specifies the simulation crystal, vacancy formation energy $E^V_f(eV)$ is shown in the second column. The third column displays the migration energies $E^V_m$, calculated from the Monomer method \cite{RAM06}. In the forth column we show the lattice parameter $a_{Fe}$(\AA). The last column displays the activation energy $E_Q(eV)$.}
\label{T4}
\begin{tabular}{l|ccccc} 
\hline 
\, Reference \, \, & \, $Fe_n$ \, & \, $E^V_f(eV)$ \, & \, $E^V_m (eV)$  \, & \, $a_{Fe}$(\AA) \, & \, $E^V_Q(eV)=E^V_f+E^V_m$\\
\hline 
\hline 
\, present work (CMS) \, & \, $Fe_{1023}$ \, & \, 1.72 \, & \, 0.68 \, & \, 2.865 \, & \, 2.40 \\
\, present work (DFT) \, & \, $Fe_{53}$ \, & \, 1.93 \, & \, 0.75 \, & \, 2.885 \, & \, 2.68\\
\, present work (DFT) \, & \, $Fe_{127}$ \, & \, 1.85 \, & \, 0.68 \, & \, 2.885 \, & \, 2.53\\
\, SIESTA+drag Method (DFT) \cite{CHU04} \, & \, $Fe_{53}$ \, & \, 2.07 \, & \, 0.67 \, & \, 2.885 \, & \, 2.74\\
\, Choudhury \cite{CHO11} \, & \, $Fe_{53}$\, & \, 2.23 \, & \, 0.67 \, & \, 2.860 \, & \, 2.90 \\
\hline
\end{tabular}
\end{center}
\end{table} 

\begin{table}[htdp]
\begin{center}
\caption{Jumps and correlated frequencies in BCC $FeCr$ using SIESTA+Monomer\cite{RAM06}. The first column denotes $C_n=Cr+V_n$ where $V_n$ means that the vacancy is $n$ nearest neighbor of the solute. Binding energies $\epsilon^V_m$ are shown in the second column. The jumps are depicted in the third column, while the forth column describes the jump frequencies ($\omega_2$,$k_{nm}$,$k_{0m}$,$k_{n0}$,$\omega_0$) \cite{OKA83} and the configurations involved in each jump. Migration energies $E^{\leftrightarrow}_m$ for direct and reversed jumps are written in the fifth and sixth column respectively. In last column results for $\epsilon^V_m$ and $E^{\leftrightarrow}_m$ from Ref. \cite{CHO11} using VASP and present calculations using SIESTA+Monomer. Binding pairs configurations are denoted as: $C_n=S+V_n$.}
\label{T5b}
\begin{tabular}{cccccc|ccc} 
\hline 
$C_n$ & $\epsilon^V_m(eV)$ & Config. & $k_{nm}$ & $E^{\rightarrow}_m(eV)$ & $E^{\leftarrow}_m(eV)$ & $\epsilon_m(eV)$ & $E^{\leftarrow}_m(eV)$ & $E^{\rightarrow}_m(eV)$ \\
\hline 
$C_{1}$ & -0.186 & 
\begin{minipage}{3.0cm} 
 \includegraphics[width=3.0cm]{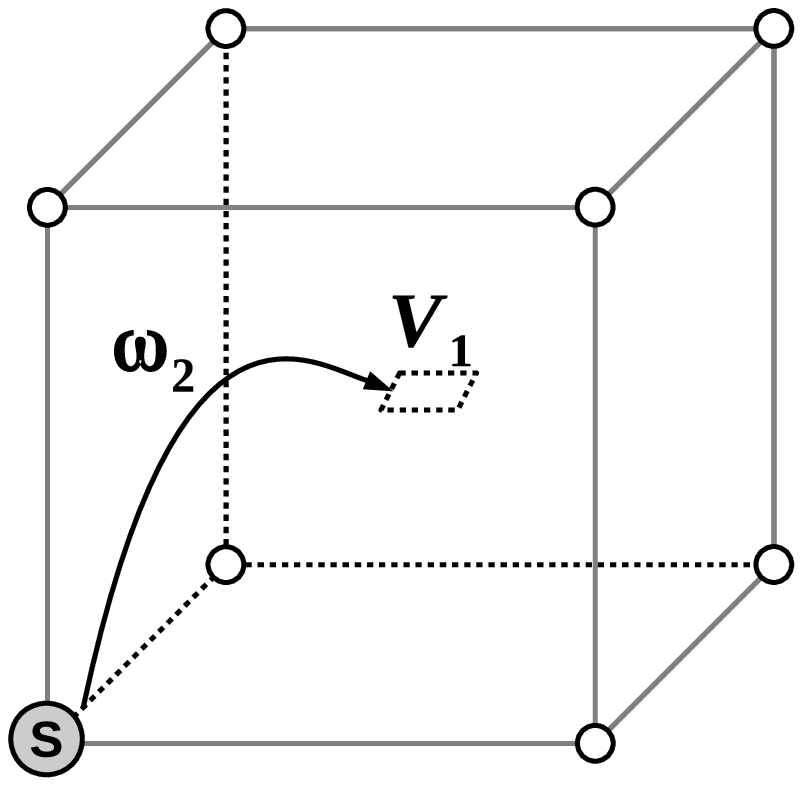} 
\end{minipage}&
$\xymatrix{
C_{1}  \ar@<0.5ex>[r]^{\omega_{2}}
& C_{1} \ar@<0.5ex>[l]^{\omega_{2}} }
$ & 0.57 & 0.57 & -0.045 & 0.58 & 0.58 \\
$C_2$ & 0.008 & 
\begin{minipage}{3.0cm} 
 \includegraphics[width=3.0cm]{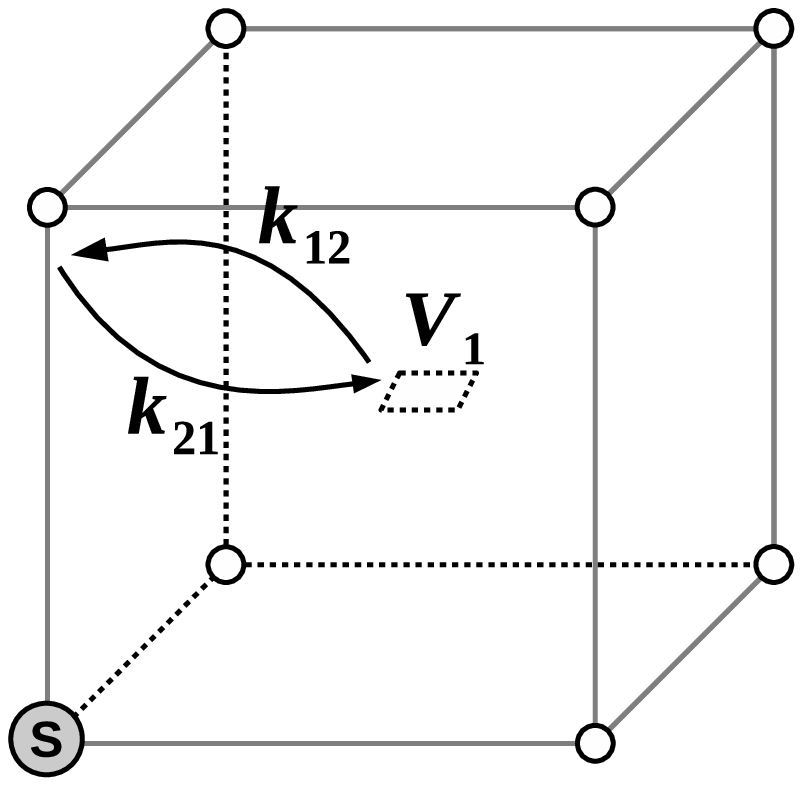} 
\end{minipage}&
$\xymatrix{
C_1  \ar@<0.5ex>[r]^{k_{21}}
& C_2 \ar@<0.5ex>[l]^{k_{12}} }
$ & 0.67 & 0.64 & -0.01 & 0.69 & 0.65 \\
$C_3$ & $0.005$ & 
\begin{minipage}{3.0cm} 
\includegraphics[width=3.0cm]{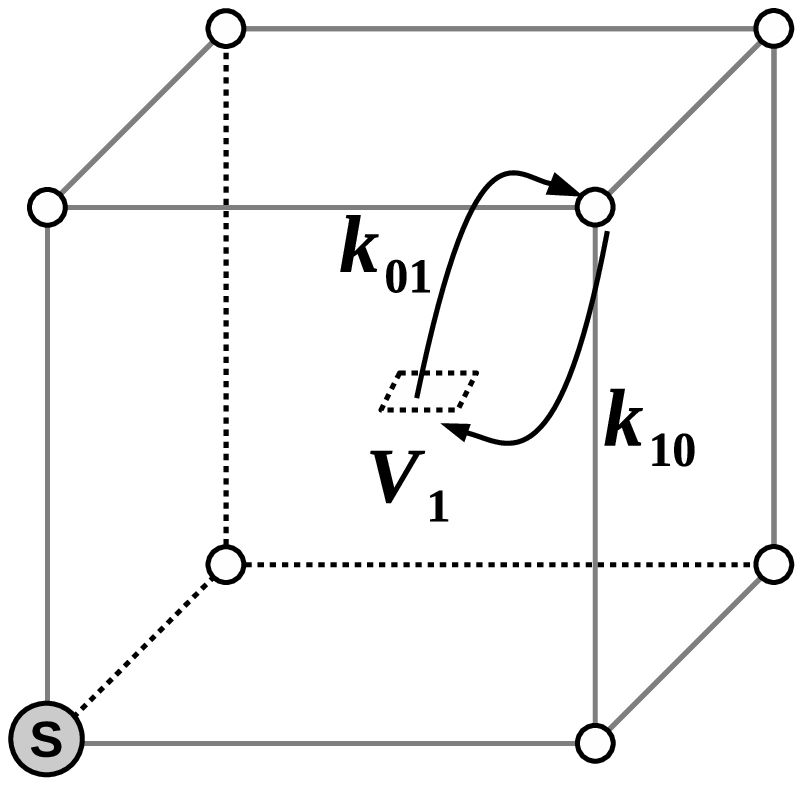} 
\end{minipage}&
$\xymatrix{
C_1  \ar@<0.5ex>[r]^{k_{01}}
& C_3 \ar@<0.5ex>[l]^{k_{10}}}
$ & 0.63 & 0.61 & -0.01 & 0.67 & 0.63 \\
$C_4$ & $0.086$ & 
\begin{minipage}{3.0cm} 
\includegraphics[width=3.0cm]{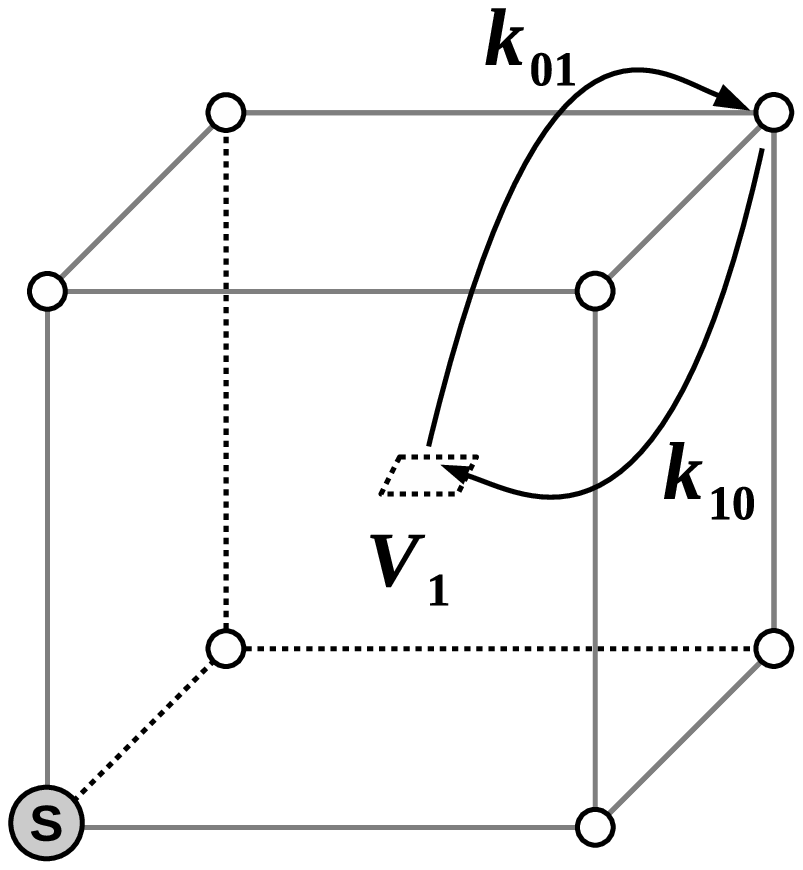} 
\end{minipage}&
$\xymatrix{
C_1  \ar@<0.5ex>[r]^{k^{\prime}_{01}}
& C_4 \ar@<0.5ex>[l]^{k^{\prime}_{10}} }$ & 0.60 & 0.59 & -0.03 & 0.64 & 0.62 \\
$C_5$ & $-0.04$ & 
\begin{minipage}{3.6cm} 
\includegraphics[width=3.6cm]{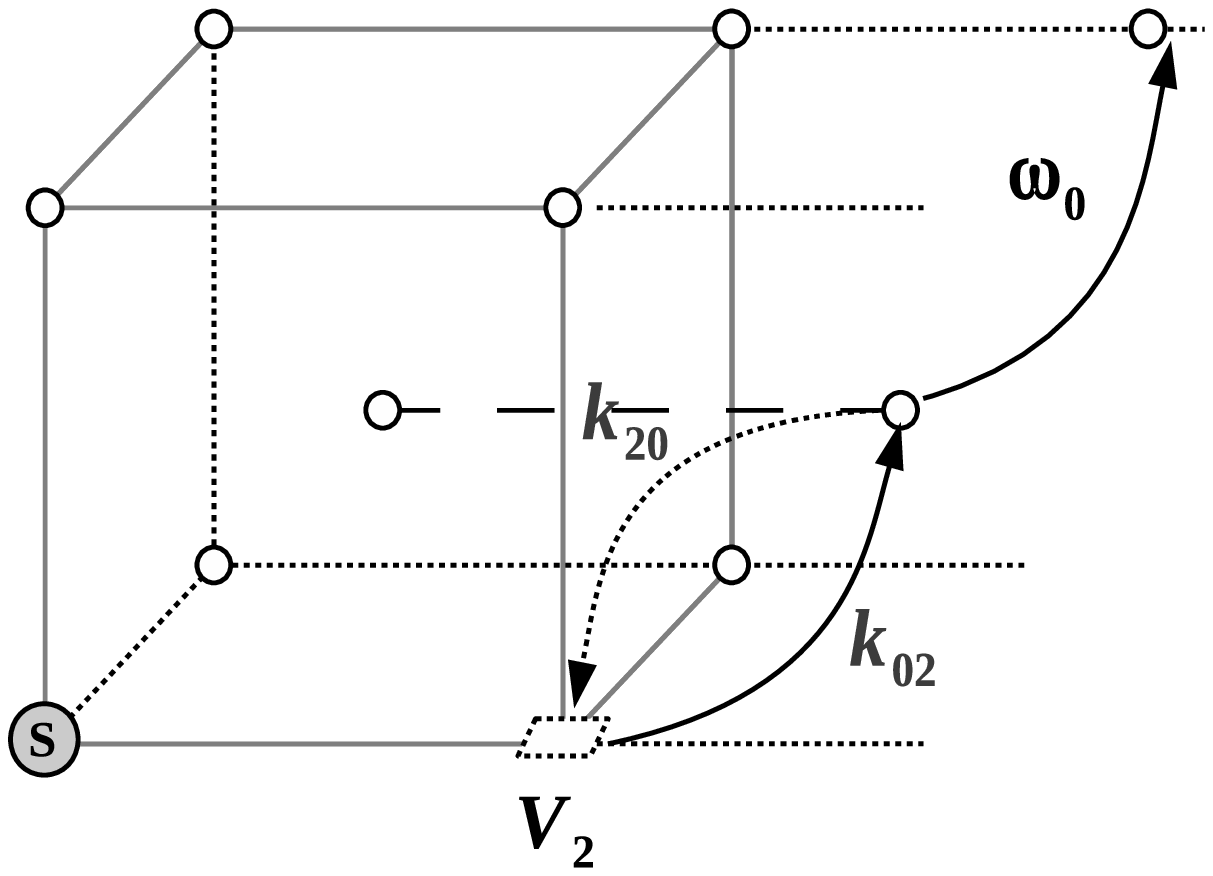} 
\end{minipage}&
$\xymatrix{
C_2  \ar@<0.5ex>[r]^{k_{02}}
& C_5 \ar@<0.5ex>[l]^{k_{20}} }$ & 0.64 & 0.66 & - & - & - \\
\hline
\end{tabular}
\end{center}
\end{table}

\clearpage
\begin{table}[htdp]
\begin{center}
\caption{Jumps and correlated frequencies in BCC $FeCr$ using CMST. The columns description is the same as in Table \ref{T5b}.}
\label{T5}
\begin{tabular}{cccccc} 
\hline 
\, $C_n=S+V_n$ \, & \, $\epsilon^V_m(eV)$ & \, Config. \, & \, $k_{nm}$ \, & \, $E^{\rightarrow}_m(eV)$ \, & \, $E^{\leftarrow}_m(eV)$ \,  \\
\hline 
\, $C_{1}$ \, & \, 0.038 \, & 
\begin{minipage}{3.0cm} 
 \includegraphics[width=3.0cm]{FIG2.eps} 
\end{minipage}&
\, $\xymatrix{
C_{1}  \ar@<0.5ex>[r]^{\omega_{2}}
& C_{1} \ar@<0.5ex>[l]^{\omega_{2}} }
$ \, & \, 0.562 \, & \, 0.562 \, \\
\, $C_2$ \, & \, 0.083 \, & 
\begin{minipage}{3.0cm} 
 \includegraphics[width=3.0cm]{FIG3.eps} 
\end{minipage}&
\, $\xymatrix{
C_1  \ar@<0.5ex>[r]^{k_{21}}
& C_2 \ar@<0.5ex>[l]^{k_{12}} }
$ \, & \, 0.670 \, & \, 0.625 \, \\
\, $C_3$ \, & \, $-0.003$ \, & 
\begin{minipage}{3.0cm} 
\includegraphics[width=3.0cm]{FIG4.eps} 
\end{minipage}&
\, $\xymatrix{
C_1  \ar@<0.5ex>[r]^{k_{01}}
& C_3 \ar@<0.5ex>[l]^{k_{10}}}
$ \, & \, 0.558 \, & \, 0.599 \, \\
\, $C_4$ \, & \, $-0.005$ \, & 
\begin{minipage}{3.0cm} 
\includegraphics[width=3.0cm]{FIG5.eps} 
\end{minipage}&
\, $\xymatrix{
C_1  \ar@<0.5ex>[r]^{k^{\prime}_{01}}
& C_4 \ar@<0.5ex>[l]^{k^{\prime}_{10}} }$ \, & \, 0.542 \, & \, 0.585 \, \\
\, $C_5$ \, & \, $0.01$ \, & 
\begin{minipage}{3.6cm} 
 \includegraphics[width=3.6cm]{FIG6.eps} 
\end{minipage}&
\, $\xymatrix{
C_2  \ar@<0.5ex>[r]^{k_{02}}
& C_5 \ar@<0.5ex>[l]^{k_{20}} }$ \, & \, 0.627 \, & \, 0.656 \, \\
\, $C_6$ \, & \, $0.01$ \, &  
\begin{minipage}{3.6cm} 
\includegraphics[width=3.6cm]{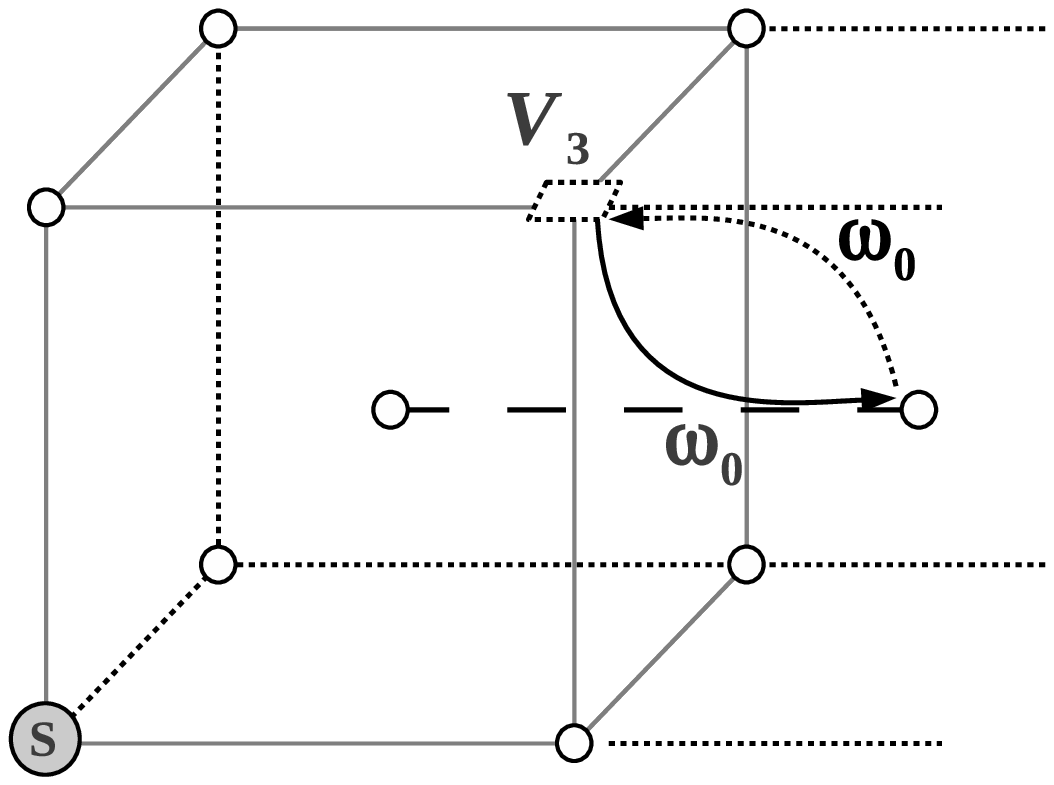} 
\end{minipage}&
\, $\xymatrix{
C_4  \ar@<0.5ex>[r]^{\omega^{\prime}_{0}}
& C_6 \ar@<0.5ex>[l]^{\omega^{\prime}_{0}} }$ \, & \, 0.605 \, & \, 0.633 \, \\
\, $C_7$ \, & \, $0.01$ \, & 
\begin{minipage}{3.6cm} 
\includegraphics[width=3.6cm]{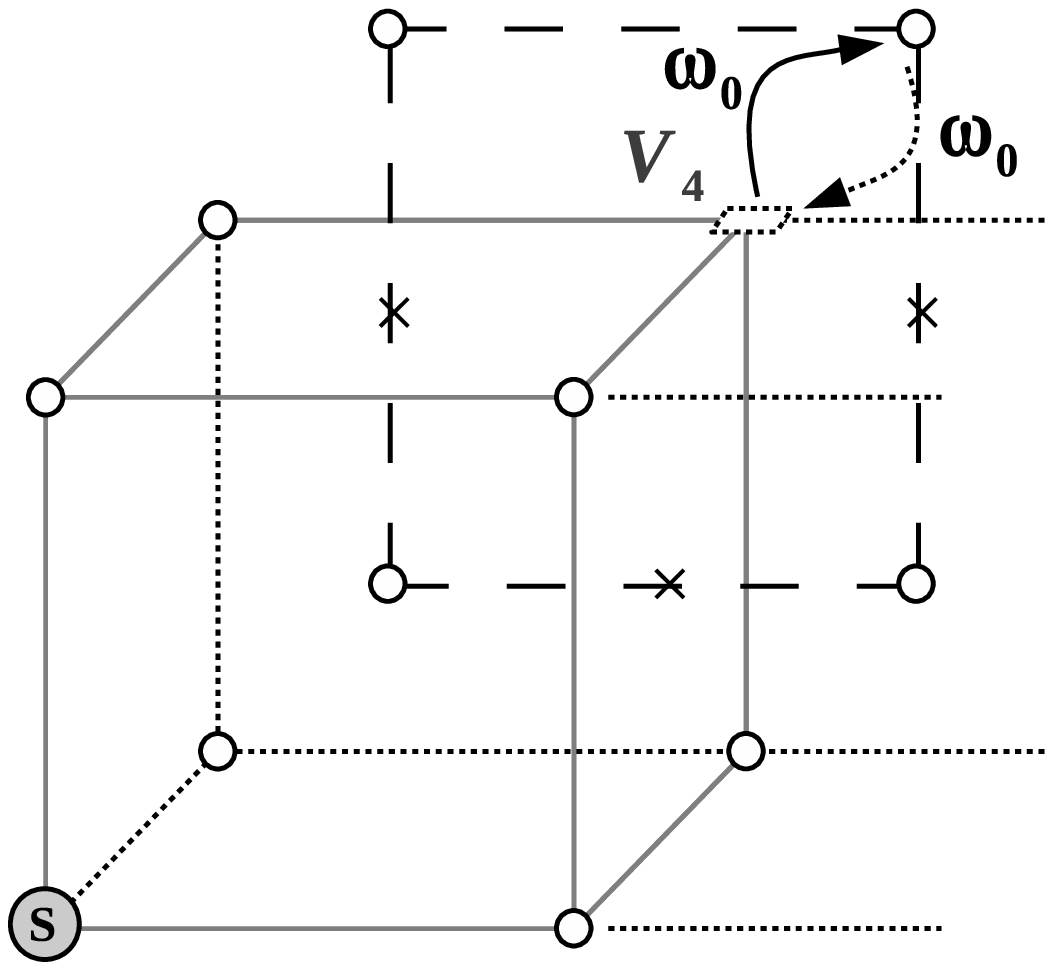} 
\end{minipage}&
\, $\xymatrix{
C_2  \ar@<0.5ex>[r]^{\omega_{0}}
& C_5 \ar@<0.5ex>[l]^{\omega_{0}} }$ \, & \, 0.639 \, & \, 0.639 \, \\
\hline
\end{tabular}
\end{center}
\end{table}

\clearpage

\noindent Binding energies are displayed in Tables \ref{T5b} and \ref{T5} respectively obtained from DFT and CMST calculations. In tables are also shown the different type of solute vacancy complex $C_n=S+V_n$ with its binding energies $\epsilon^V_m$, together with the possibles configurations and jumps that involve the corresponding $C_n=S+V_n$ complex with the corresponding jump frequencies. Vacancy migration barriers are shown for the direct as well as for the reverse jumps. Results from CMST shown a weak repulsive/attractive energy interaction, $\epsilon^V_m$, between the vacancy and solute in all vacancy-solute pairs configurations, while from DFT calculations shown in Table \ref{T5b}, reveal a strong binding energy at first nearest neighbor $Cr+V_1$ pair and a weak repulsive/attractive interaction for the rest of the pairs. The binding energies in Ref. \cite{CHO11} agrees with those obtained from CMST in present calculations. We note that, the biggest difference is observed in the value of the vacancy formation energy, which is involved in the calculation of the full set of $L$-coefficients and therefore in the diffusion coefficients.

Table \ref{T5} also shows that, the migration barriers and corresponding frequencies are in well agreement, using either DFT or CMST calculations excepting for the case of $E^{\leftrightarrow}_m$ related to $k_{01}$ in transitions $C_1 \rightarrow C_3$ and $C_1 \rightarrow C_4$. 

In order to obtain the jump frequencies, we assume that the jumps are thermally activated and we use equations (\ref{nu0T}) and (\ref{nuT}). Expression (\ref{nu0T}), is written in terms of $E^{\rightarrow}_m$ which are reported in Table \ref{T5}. For the pre-factor in (\ref{nu0T}), we use a constant attempt frequency $\nu _{0}=5\times 10^{12}Hz$, taken from Ref. \cite{CHO11} for pure $Fe$. We also use, in terms of the Wert model \cite{WER65}, a temperature dependent attempt frequency given by (\ref{nuT}). Also in Table \ref{T5}, the migration barriers and the corresponding rate frequency for each jump are shown. Table \ref{T8} presents the calculated frequencies at two different temperatures and for both a constant and a temperature dependent pre-exponential factor $\nu_0(T)$ from (\ref{nuT}) using CMST and DFT calculation.
\begin{table}[htdp]
\begin{center}
\caption{Vacancy jump frequencies rate $\omega_0$, $\omega_2$ and $k_{nm}$ calculated with a constant and a temperature dependent attempt frequency, at two  different temperatures in $FeCr$ alloy using CMST and DFT calculations.}
\label{T8}
\begin{tabular}{c|cc|cc|cc} 
\hline 
\, & \,\,\,\,\,\,\,\,\,\, $\nu_0=5\times 10^{12}Hz$ \, &CMST& \, $\nu_0=1/\beta h$\, &CMST& \, $\nu_0=1/\beta h$ DFT&using SIESTA
\tabularnewline 
\hline 
\, & \, $T_1=300K$\, & \, $T_2=1250K$\, & \, $T_1=300K$\, & \, $T_2=1250K$\, & \, $T_1=300K$\, & \, $T_2=1250K$\,  
\tabularnewline 
$\omega_i$ \, & & & & & & 
\tabularnewline 
\hline 
$\omega_0$ \, & \, $18.84\times 10^{0}$ \, &\, $9.06\times 10^{9}$ \, & \, $23.55\times 10^{0}$ \, &\, $4.72\times 10^{10}$ \, & \, $34.68\times 10^{0}$ \, &\, $5.18\times 10^{10}$  
\tabularnewline 
$\omega_2$ \, &\, $18.09\times 10^{2}$ \, & \, $2.71\times 10^{10}$ \, & \, $22.62\times 10^{2}$ \, &\, $1.41\times 10^{11}$ \, & \, $16.60\times 10^{2}$ \, &\, $1.31\times 10^{11}$ \, 
\tabularnewline  
$k_{21}$ \, & \, $27.74\times 10^{0}$ \, & \, $9.94\times 10^{9}$ \, & \, $34.67\times 10^{0}$ \, &\, $5.18\times 10^{10}$ \, & \, $34.67\times 10^{0}$ \, &\, $5.18\times 10^{10}$ \,
\tabularnewline 
$k_{12}$ \, & \, $15.81\times 10^{1}$ \, & \, $1.51\times 10^{10}$ \, & \, $19.77\times 10^{1}$ \, &\, $7.87\times 10^{10}$ \, & \, $11.07\times 10^{1}$ \, &\, $6.84\times 10^{10}$ \, 
\tabularnewline 
$k_{01}$ \, & \, $21.12\times 10^{2}$ \, & \, $2.81\times 10^{10}$ \, & \, $26.40\times 10^{2}$ \, &\, $1.47\times 10^{11}$ \, & \, $16.29\times 10^{1}$ \, &\, $7.51\times 10^{10}$ \, 
\tabularnewline 
$k_{02}$ \, & \, $14.64\times 10^{1}$ \, & \, $1.48\times 10^{10}$ \, & \, $18.30\times 10^{1}$ \, &\, $7.72\times 10^{10}$ \, & \, $35.32\times 10^{1}$ \, &\, $9.04\times 10^{10}$ \, 
\tabularnewline 
\hline
\end{tabular}
\end{center}
\end{table}

\begin{figure}[h]
\begin{center}
\includegraphics[angle=-90,width=10.0cm]{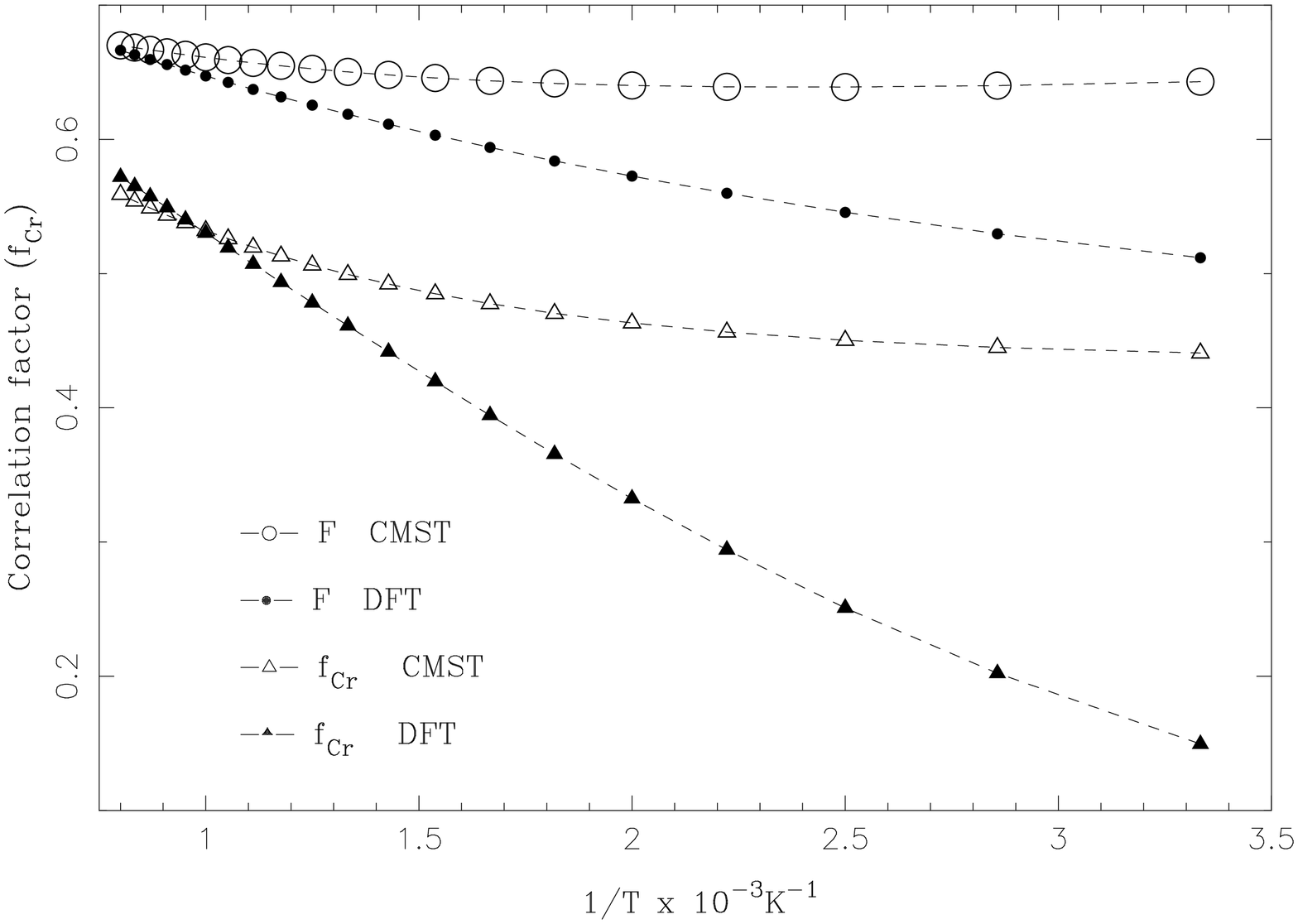}
\vspace{1.3cm}
\caption{Solute correlation factor $f_{Cr^{\star}}$ in the $FeCr$ system as a function of the temperature in the second shell approximation. The $F$ factor is denoted with up triangles.}
\label{FIG16}
\end{center}
\end{figure}

We calculate the solute correlation factors $f_{Cr}$ from (\ref{fCr}) in term of the above calculated frequencies in Table \ref{T8} using CMST, and also from frequencies obtained from DFT calculations using SIESTA+Monomer. The solute-correlation factor ($f_{Cr}$) with $T$, calculated from (\ref{fCr}), is shown in Table \ref{T9} and Figure \ref{FIG16}. The factor $F$ obtained from equation (\ref{Q11}) is also shown. Also, table \ref{T9}, resumes the jump frequencies ratios calculated from CMST according to the BCC six-frequency model of solute-vacancy interaction for a constant pre-exponential frequency factor.

\begin{table}[h]
\begin{center}
\caption{Solute correlated factors for $FeCr$ at different temperatures from CMST calculations, in the second shell approximation for a constant $\nu_0=5\times10^{12}$Hz value. First and second columns display respectively the alloy and the temperature range. Results of solute correlated factor $f_{Cr}$ are shown in column three. The last tree columns describe the jump frequency ratios of the solute$-$vacancy interaction.}
\label{T9}
\begin{tabular}{crccccc}
\hline
\hline
\, Alloy \, & \, $T/K$ \, & \, $f_{Cr}$ \, & \, $k_{01}/\omega_0$ \, & \, $k_{02}/\omega_0$ \, & \, $k_{21}/k_{12}$ \, & \, $k_{12}/\omega_{0}$ \, \tabularnewline
\hline 
\, $FeCr$ \, & \, 300 \, & \, 0.643 \, & \, 112.09 \, & \, 7.77 \, & \, 0.18 \, & \, 8.39 \, \tabularnewline 
\,  \, & \, 350 \, & \, 0.640 \, & \, 57.12 \, & \, 5.80 \, & \, 0.22 \, & \, 6.19 \, \tabularnewline 
\,  \, & \, 400 \, & \, 0.639 \, & \, 34.45 \, & \, 4.65 \, & \, 0.27 \, & \, 4.93 \, \tabularnewline 
\,  \, & \, 450 \, & \, 0.639 \, & \, 23.25 \, & \, 3.92 \, & \, 0.31 \, & \, 4.13 \, \tabularnewline 
\,  \, & \, 500 \, & \, 0.640 \, & \, 16.97 \, & \, 3.42 \, & \, 0.35 \, & \, 3.58 \, \tabularnewline 
\,  \, & \, 550 \, & \, 0.642 \, & \, 13.12 \, & \, 3.06 \, & \, 0.39 \, & \, 3.19 \, \tabularnewline 
\,  \, & \, 600 \, & \, 0.644 \, & \, 10.59 \, & \, 2.79 \, & \, 0.42 \, & \, 2.89 \, \tabularnewline 
\,  \, & \, 650 \, & \, 0.646 \, & \, 8.83 \, & \, 2.58\, & \, 0.44 \, & \, 2.67 \, \tabularnewline 
\,  \, & \, 700 \, & \, 0.648 \, & \, 7.56 \, & \, 2.41 \, & \, 0.47 \, & \, 2.49 \, \tabularnewline 
\,  \, & \, 750 \, & \, 0.650 \, & \, 6.60 \, & \, 2.27 \, & \, 0.50 \, & \, 2.34 \, \tabularnewline 
\,  \, & \, 800 \, & \, 0.653 \, & \, 5.87 \, & \, 2.16 \, & \, 0.52 \, & \, 2.22 \, \tabularnewline 
\,  \, & \, 850 \, & \, 0.655 \, & \, 5.29 \, & \, 2.06 \, & \, 0.54 \, & \, 2.12 \, \tabularnewline 
\,  \, & \, 900 \, & \, 0.657 \, & \, 4.82 \, & \, 1.98 \, & \, 0.56 \, & \, 2.03 \, \tabularnewline 
\,  \, & \, 950 \, & \, 0.659 \, & \, 4.34 \, & \, 1.91 \, & \, 0.58 \, & \, 1.96 \, \tabularnewline 
\,  \, & \, 1000 \, & \, 0.661 \, & \, 4.12 \, & \, 1.85 \, & \, 0.59 \, & \, 1.89 \, \tabularnewline 
\,  \, & \, 1050 \, & \, 0.663 \, & \, 3.85 \, & \, 1.79 \, & \, 0.61 \, & \, 1.84 \, \tabularnewline 
\,  \, & \, 1100 \, & \, 0.664 \, & \, 3.62 \, & \, 1.75 \, & \, 0.62 \, & \, 1.78 \, \tabularnewline 
\,  \, & \, 1150 \, & \, 0.667 \, & \, 3.43 \, & \, 1.71 \, & \, 0.64 \, & \, 1.74 \, \tabularnewline 
\,  \, & \, 1200 \, & \, 0.668 \, & \, 3.25 \, & \, 1.67 \, & \, 0.65 \, & \, 1.70 \, \tabularnewline 
\,  \, & \, 1250 \, & \, 0.670 \, & \, 3.10 \, & \, 1.64 \, & \, 0.66 \, & \, 1.67 \, \tabularnewline 
\hline 
\end{tabular}
\end{center}
\end{table}

\begin{figure}[h]
\begin{center}
\includegraphics[angle=-90,width=10.0cm]{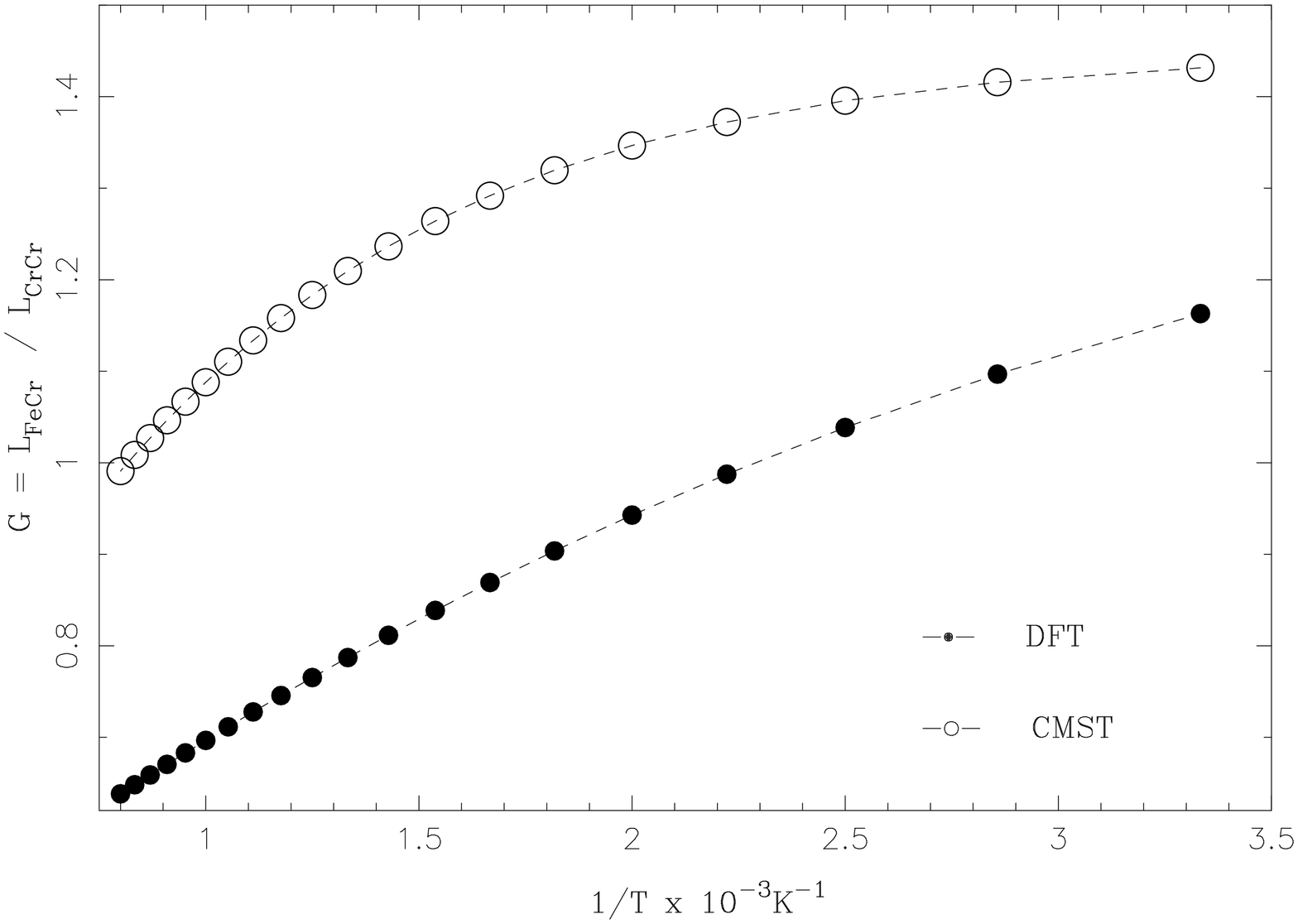}
\vspace{1.3cm}
\caption{Ratio of the vacancy-Onsager coefficients of $Cr$ in $Fe$ calculated from expressions $L_{FeCr}$ and $L_{CrCr}$ in Ref. \cite{BAR07} \textit{vs} $1/T$.}
\label{FIG17}
\end{center}
\end{figure}

The Onsager and Diffusion coefficients were calculated assuming a solute mole fraction of $C_{Cr}=0.001$, which corresponds to $n_{Cr}=NC_{Cr}=8.5\times 10^{19}cm^{-3} \, atoms/cm^3$ and $N=8.5\times 10^{22}cm^{-3} \, atoms/cm^3$ is the number of atoms/volume. Once calculated $L_{FeCr}$ and $L_{CrCr}$, and following the reasoning in Ref. \cite{CHO11}, we also calculate the vacancy wind coefficient $G=L_{FeCr}/L_{CrCr}=-(1+L_{VCr}/L_{CrCr})$. The results are presented in Figure \ref{FIG17} for both DFT and CMST calculations. We see that $G>-1$ in all the temperature range considered, showing that the vacancy drag mechanism is unlikely to occur using the present description, in agreement with results in Refs. \cite{CHO11} and \cite{MES13}.

\begin{figure}[h]
\begin{center}
\includegraphics[angle=-90,width=10.0cm]{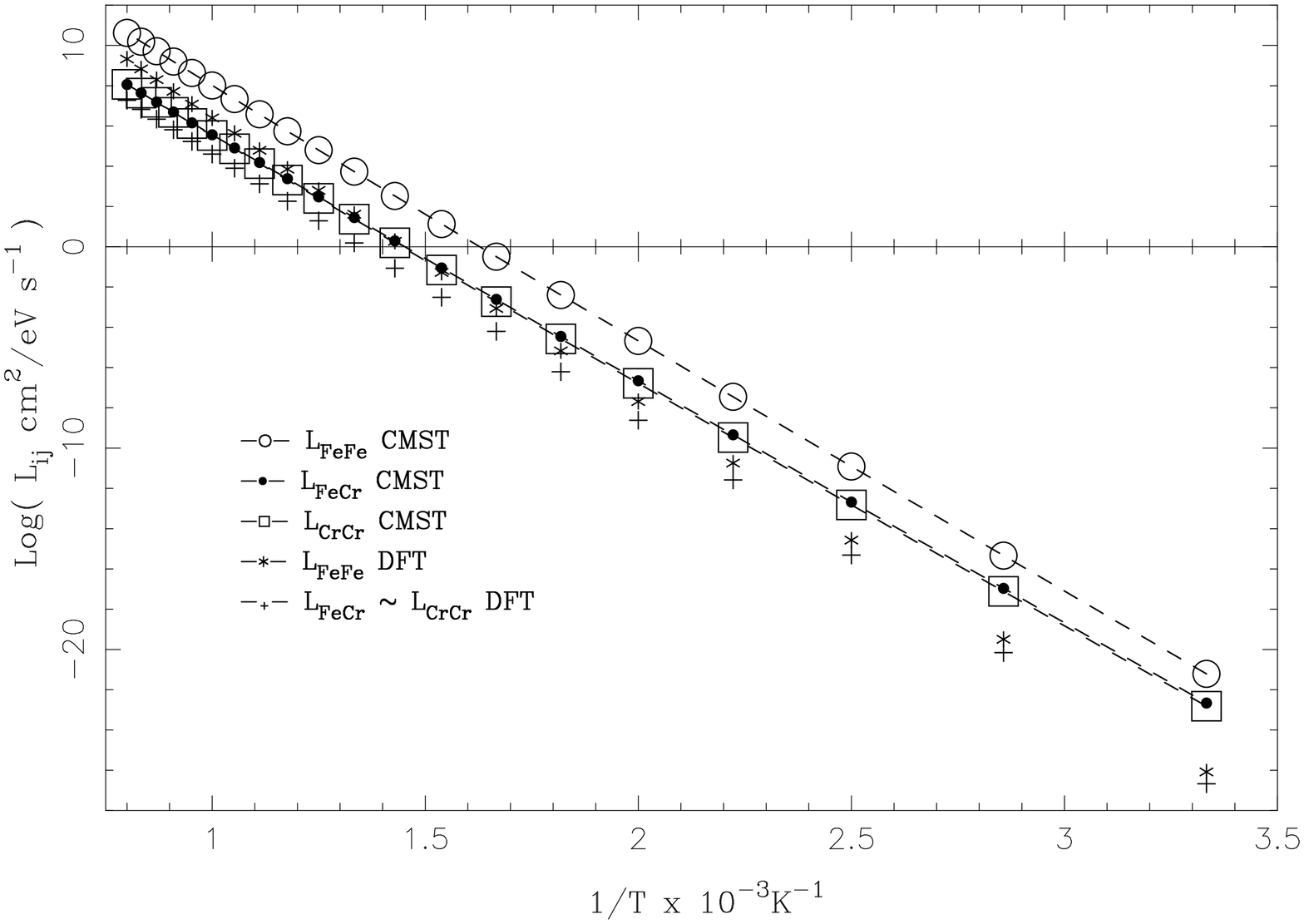}
\vspace{1.3cm}
\caption{Vacancy-Onsager coefficients \textit{vs} $1/T$ for the $FeCr$ system. Squares denote $L_{CrCr}$, empty circles denote $L_{FeFe}$ while $L_{FeCr}$ is described with filled circles.}
\label{FIG18}
\end{center}
\end{figure}

\begin{figure}[h]
\begin{center}
\includegraphics[angle=-90,width=10.0cm]{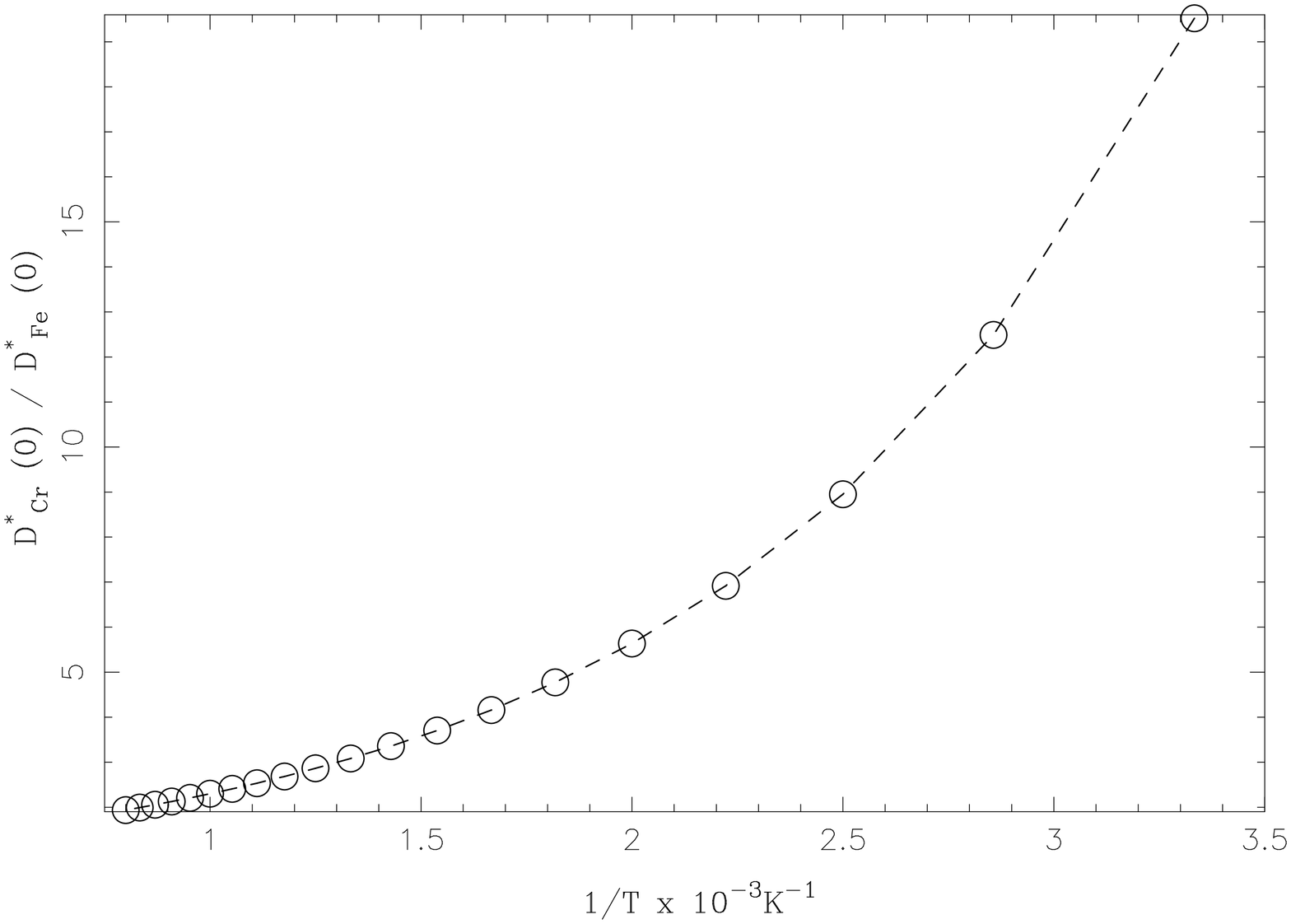}
\vspace{1.3cm}
\caption{Ratio of the tracer diffusion coefficient $D^{\star}_{Cr}/D^{\star}_{Fe}$ in $FeCr$ \textit{vs} $1/T$ from CMST calculations.}
\label{FIG19}
\end{center}
\end{figure}

\begin{figure}[h]
\begin{center}
\includegraphics[angle=-90,width=8.5cm]{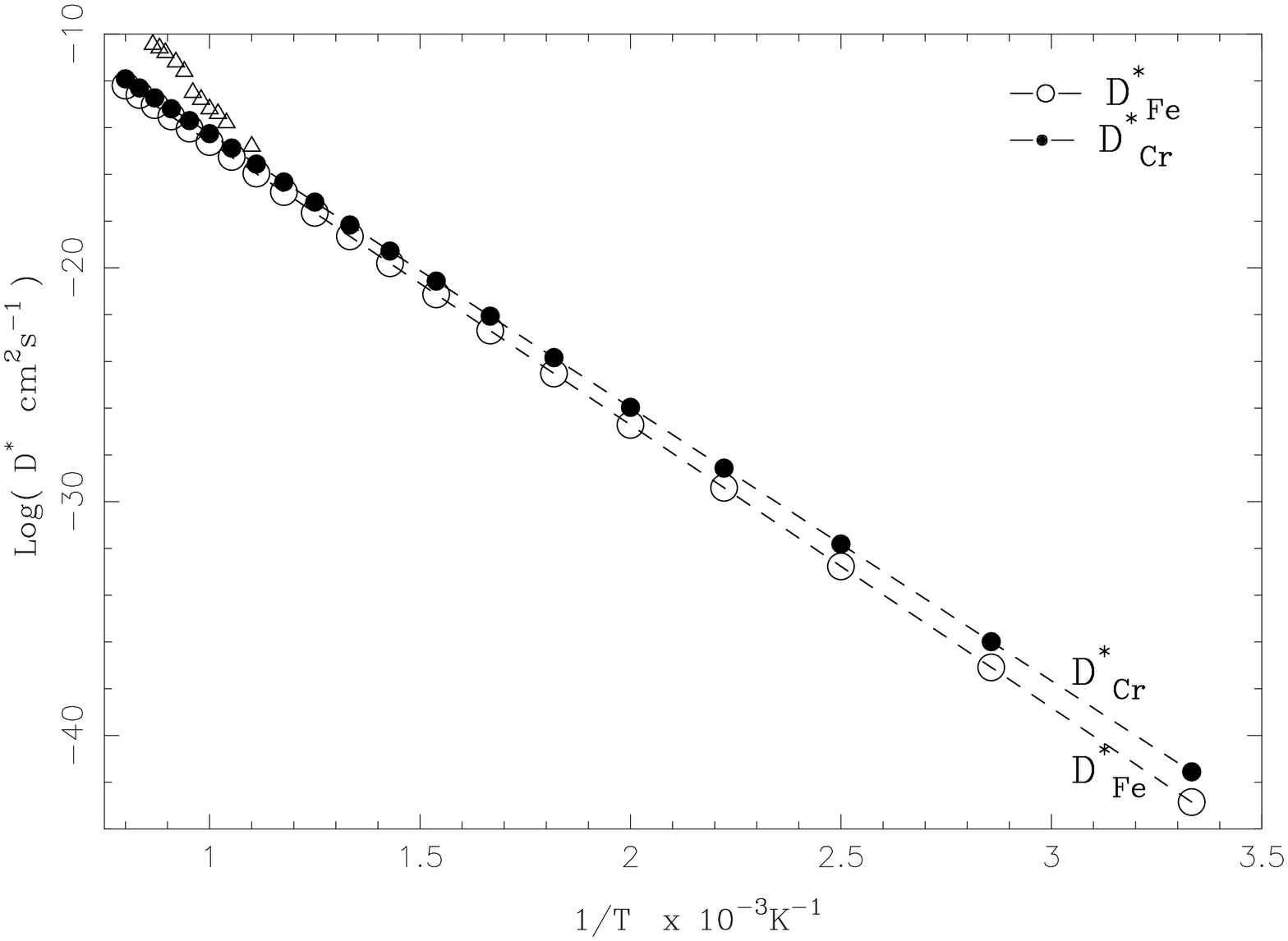}\hspace{0.5cm}\includegraphics[angle=-90,width=6.0cm]{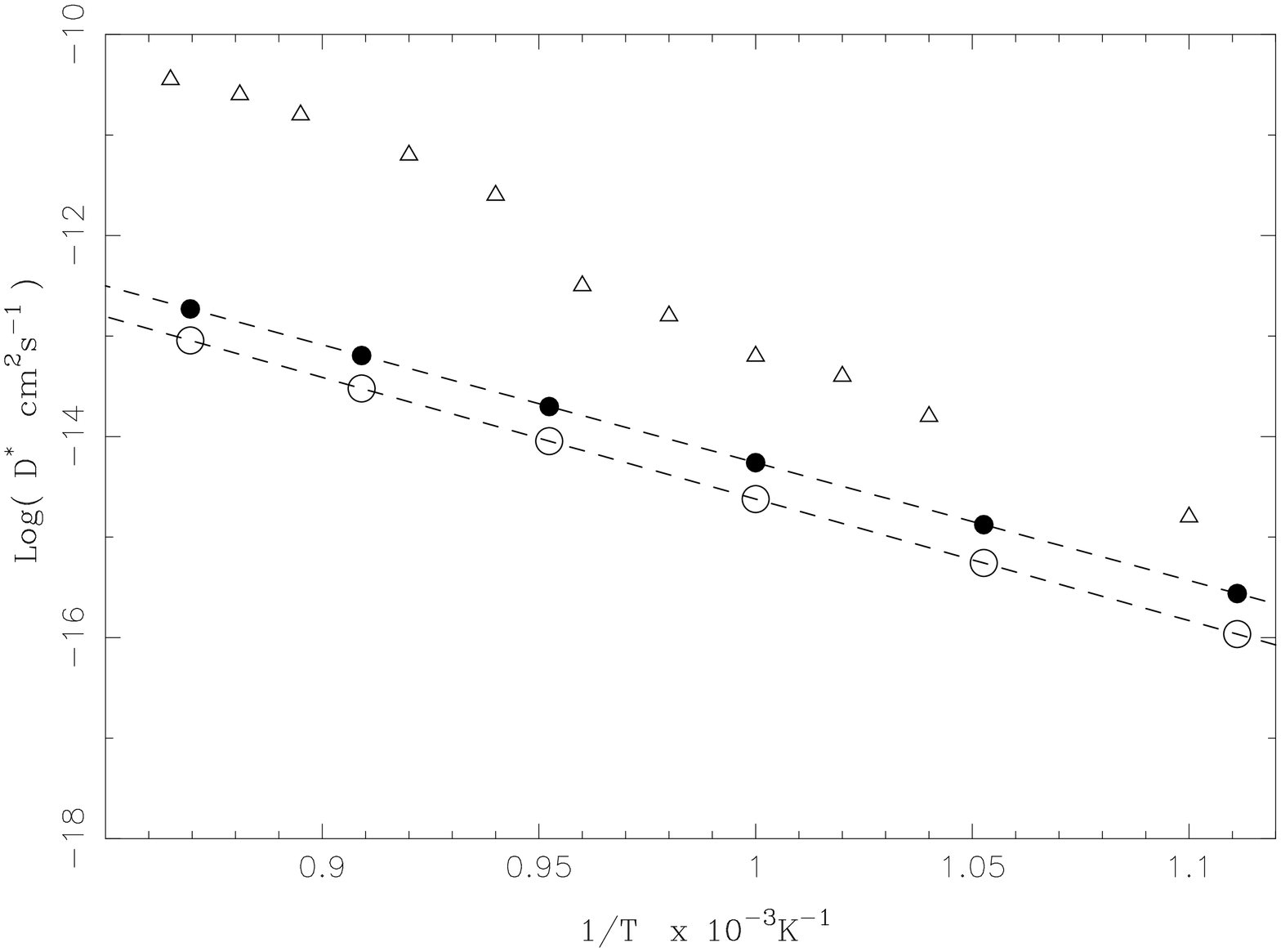}
\caption{Tracer diffusion coefficients of $Cr$ ($D^{\star}_{Cr}$ in filled black circles) and $Fe$ ($D^{\star}_{Fe}$ in open circles) in the alloy, for a constant attempt frequency $\nu_0=5\times10^{12}$Hz. Available experimental data, for the $Cr$ diffusion coefficient in the alloy, are displayed with open triangles \cite{CHO11}.}
\label{FIG20a}
\end{center}
\end{figure}

\begin{figure}[h]
\begin{center}
\includegraphics[angle=-90,width=8.0cm]{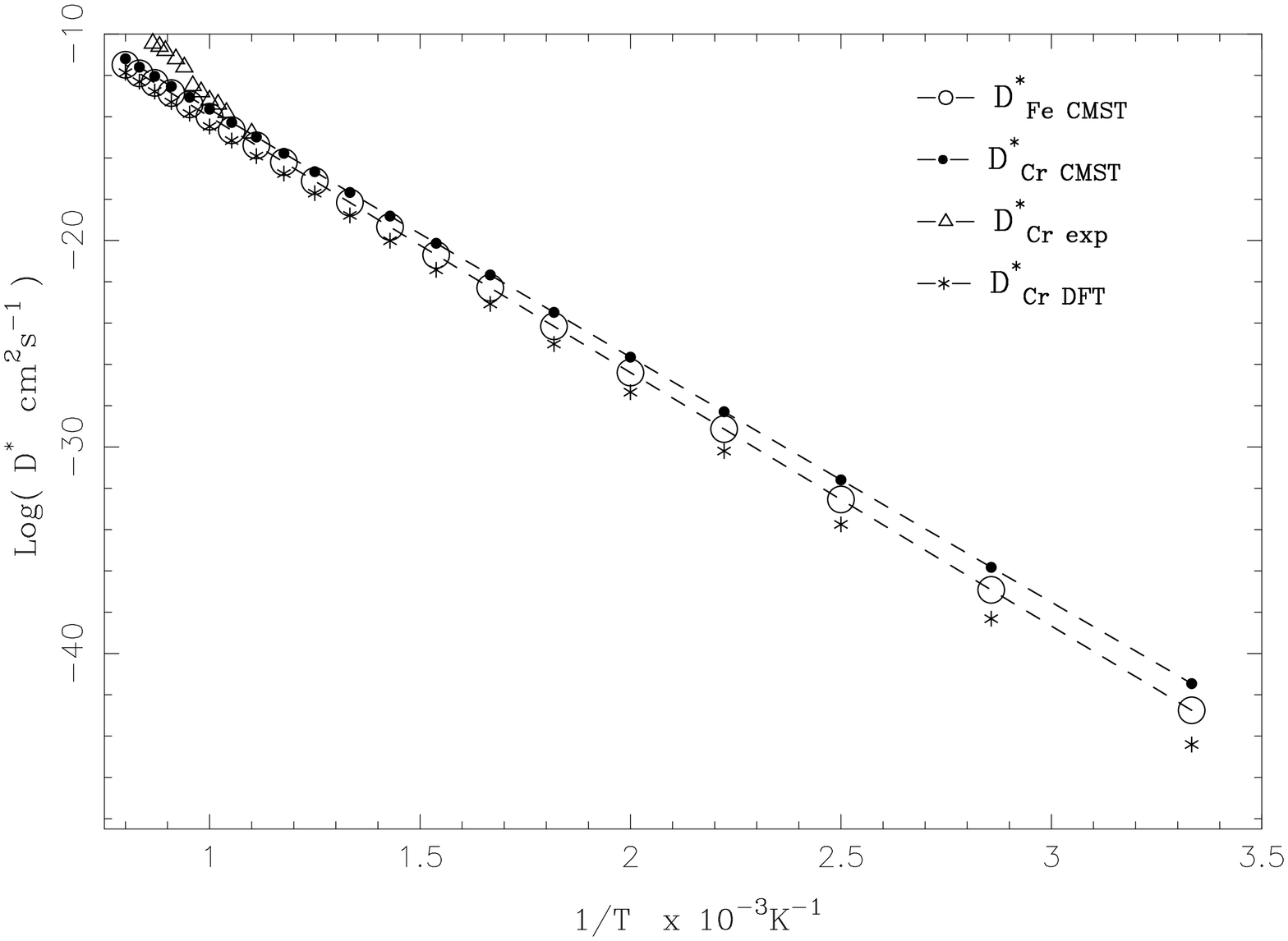}\hspace{0.7cm}\includegraphics[angle=-90,width=6.0cm]{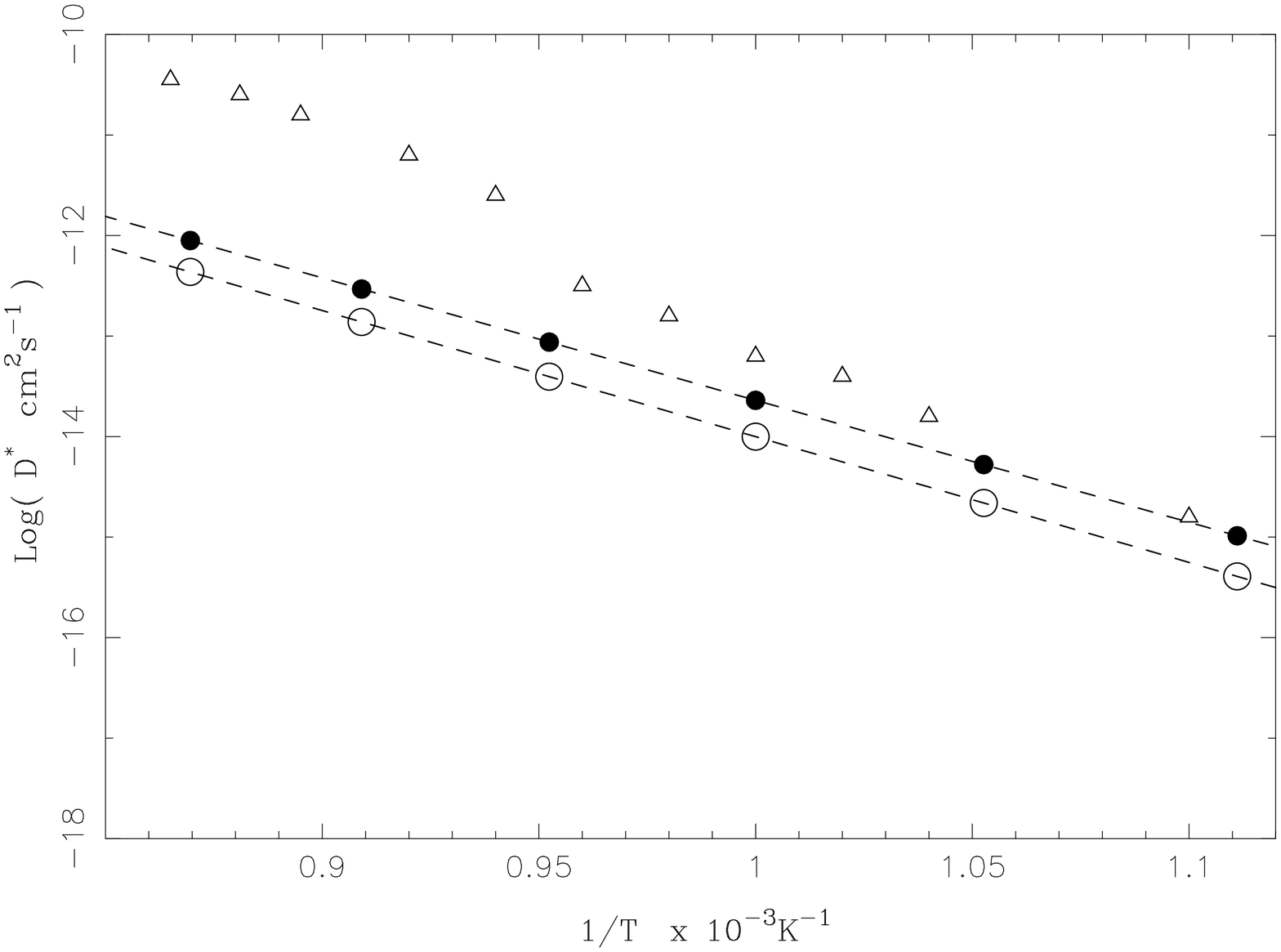}
\vspace{1.cm}
\caption{Tracer diffusion coefficients of $Cr$ ($D^{\star}_{Cr}$ in filled black circles) and $Fe$ ($D^{\star}_{Fe}$ in open circles) in the alloy, for a temperature dependent attempt frequency $\nu_0(T)=k_BT/h$ in (\ref{nuT}). Available experimental data, for the $Cr$ diffusion coefficient in the alloy, are displayed with open triangles \cite{CHO11}.}
\label{FIG20c}
\end{center}
\end{figure}

The full set of $L$-coefficients are displayed in Fig. \ref{FIG18} against $1/T$, $T$ in $^{\circ}K$. We see that the $L$-coefficients follow an Arrhenius behavior, which implies a linear relation between the logarithm of $L$-coefficients against the inverse of the temperature (see Fig. \ref{FIG18}).
Now, we are in position to obtain the diffusion coefficients $D^{\star}_{Fe}(0)$, $D^{\star}_{Cr}(0)$. First, we present the ratio of calculated tracer diffusion coefficients $D^{\star}_{Cr}/D^{\star}_{Fe}$ as a function of the inverse of the temperature in Figure \ref{FIG19}. 
The calculated $D^{\star}_{Fe}$ and $D^{\star}_{Cr}$, using equations (\ref{DFe0}) and (\ref{DCr}), are shown in Figures \ref{FIG20a} and \ref{FIG20c} for a constant and a temperature dependent attempt frequency $\nu_0$ respectively. It is import to perform a comparison between theoretical results obtained in present work with reliable experimental data. In Figs. \ref{FIG20a} and \ref{FIG20c} experimental data of the solute diffusion coefficient are plotted with open triangles on the temperature range of $T=[860-1200]^{\circ}C$ \cite{CHO11}. At right we show a magnifier of the temperature range of experimental data are. As we can see an excellent agreement between calculated solute diffusion coefficient $D^{\star}_{Cr}$ using a CMST and experimental values occurs specially below to the $Fe$-solvent melting temperature $T_m(Fe)=1043^{\circ}C$ for a temperature dependent attempt frequency $\nu_0(T)$. Although in Choudhury et al. theoretical calculations describes the non-Arrhenius behavior of $Cr$ and $Fe$ diffusion in the alloy, their ab-initio calculations underestimates the solvent and solute-diffusion coefficient values by more than four orders of magnitude in the temperature range where experimental data have been considered. Also in Fig. \ref{FIG20c} in stars symbols, we show the tracer solute diffusion coefficient $D^{\star}_{Cr}$ from present DFT calculations.

\clearpage

In Ref \cite{CHO11} the activation energy, $Q=E^{\rightarrow}_m+E_F^V$, is corrected in considering the phase transition from a paramagnetic to a ferromagnetic state \cite{CRA71} at temperatures below $1043 K$. In the paramagnetic state $D^{\star}_{Fe}$ for pure iron follows an Arrhenius relationship, but below the Curie temperature the $D^{\star}_{Fe}$ deviates from the Arrhenius behavior. In the ferromagnetic state the correct activation energy for vacancy diffusion is a function of the spontaneous magnetization which is expressed as \cite{RUC76},
\begin{equation}
\Delta Q^F(T)=\Delta Q^P(T)(1+\alpha s(T)^2)
\label{DQcor}
\end{equation}
where $Q^P$ is the activation energy in the paramagnetic state, $s(T)$ is the ratio of the spontaneous magnetization of pure iron at a given temperature to the spontaneous magnetization at $0K$ \cite{CRA71}. The constant $\alpha $ takes into account the change in formation and binding energies due to magnetic transformation. The value of a is obtained empirically in Refs. \cite{NIT05,TAK07} for pure Fe and Fe-Cr system to be 0.156 and 0.133 respectively. We do not correct here the activation energy.

\begin{figure}[h]
\begin{center}
\includegraphics[width=8.0cm]{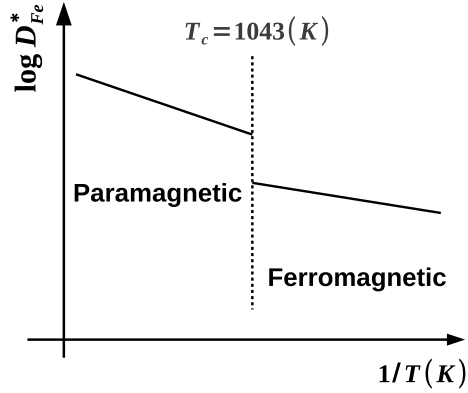}
\caption{Spontaneous magnetization in Fe.}
\end{center}
\end{figure}

In summary, present calculations for diluted alloys, using both CMST and DFT, confirm that a vacancy drag mechanism is unlikely to occur in $FeCr$ diluted alloys as previously shown by Choudhury \cite{CHO11}. On the other hand, present calculations are in better agreement with experimental $D^{\star}_{Fe}$ and $D^{\star}_{Cr}$ values, than DFT calculations in Ref. \cite{CHO11}. Although the authors describes correctly the non Arrhenius behavior at the melting temperature, $T_m(Fe)$, their values of the theoretical diffusion coefficients are 4 orders of magnitude below the experimental data. We assume that the improvement in our calculations, in comparison with calculations in Ref. \cite{CHO11}, mainly depends on the second nearest neighbor binding description here employed confirmed by our oun DFT calculations using SIESTA+Monomer. Recently a study performed by Garnier et al \cite{GAR13}, shows that a third shell approach may lead to a vacancy drag mechanism at low temperatures in BCC structures. This fact deserves that our attention and, in a future work we will report new calculations using this new model to obtain convergent results independently of the approximation employed.  

\subsection{Diffusion driven by interstitial migration}

\vspace{0.5cm}
For diffusion mediated by interstitial migration in bulk, we use the final expressions for $L-$coefficients in the second shell approximation developed by Barbe and Nastar \cite{BAR07}. In Table \ref{TIf}, we present our results for the interstitial formation energy ($E^I_f$) in pure $\alpha Fe$ calculated as,
\begin{equation}
 E^I_f=E(N+1)-E_c-E(N),
\end{equation}
As for vacancies, $E(N)$ corresponds to the energy of relaxed lattice containing $N$ atoms, $E(N+1)$ is the energy of the defective system containing one single interstitial, and $E_c$ is the $Fe$ atom cohesion energy. Migration barrier of the Interstitial in perfect lattice ($E^I_m$), is also calculated with the Monomer method \cite{RAM06}. Here, as for vacancies, a complex $Cr+I_k$, of subsitutional $Cr$ nearest neighbor of a pure $Fe$ dumb-bell $I_k$, is formed. Mixed dumb-bell configurations, which leads to solute migration, are also considered. The binding energy between $Cr$ and $I_k$ is calculated respectively as,
\begin{equation}
\epsilon ^I_m=\left\{ E(N-1,Cr+I_k) + E(N)\right\} - \left\{E(N-1,Cr)+E(N+1,I)\right\} ,
\label{EbnnI}
\end{equation}
where $E(N+1,I)$ and  $E(N-1,Cr)$ are the energies of a crystallite containing ($N-1$) atoms of solvent $Fe$ plus one $Fe$ interstitial $I$, and one solute atom of $Cr$ respectively, while $E(N-1,C_n=Cr+I_k)$ is the energy of the crystallite containing ($N-1$) atoms of $Fe$ plus one solute-interstitial complex $C_n=Cr+I_k$. With the sign convention used here $\epsilon ^I_m < 0$ means attractive solute-interstitial interaction, and $\epsilon ^I_m > 0$ indicates repulsion. As before, we computete the migration energies $E^{\leftrightarrow}_m$ using the Monomer Method \cite{RAM06}. CMST and DFT calculations show that the minimum energy configurations are so that $E(C^{\perp}_0) \simeq E(C^{m}_0)<E(C^{\parallel}_0)$, which implies that $C^{\perp }_0$ corresponds to the absolute minimum configuration.  
\clearpage
\begin{table}[ht] 
\begin{center}
\caption{Energies and lattice parameters for the pure BCC $Fe$ lattice. The first column specifies the simulation crystal, Interstitial formation energy $E^I_f(eV)$ is shown in the second column. The third column displays the migration energies $E^I_m$ respectively for $Fe$ dumb-bell roto-translation/pure translation, calculated from the Monomer method \cite{RAM06}. In the forth column we show the lattice parameter $a_{Fe}$(\AA).}
\label{TIf}
\begin{tabular}{l|cccc} 
\hline 
\, Reference \, \, & \, $Fe_n$ \, & \, $E^I_f(eV)$ \, & \, $E^I_m =E^I_Q (eV)$  \, & \, $a_{Fe}$(\AA) \,\\
\hline 
\hline 
\, present work using CMST \, & \, $Fe_{1025}$ \, & \, 3.53 \, & \, 0.31/0.85 \, & \, 2.885 \, \\
\, Choudhury \cite{CHO11} using VASP \, & \, $Fe_{55}$\, & \, $\sim 4$ \, & \, 0.35/0.84 \, & \, 2.860 \, \\
\hline
\end{tabular}
\end{center}
\end{table} 

\begin{figure}[h]
\begin{center}
\hspace{-.8cm}\includegraphics[width=6.cm]{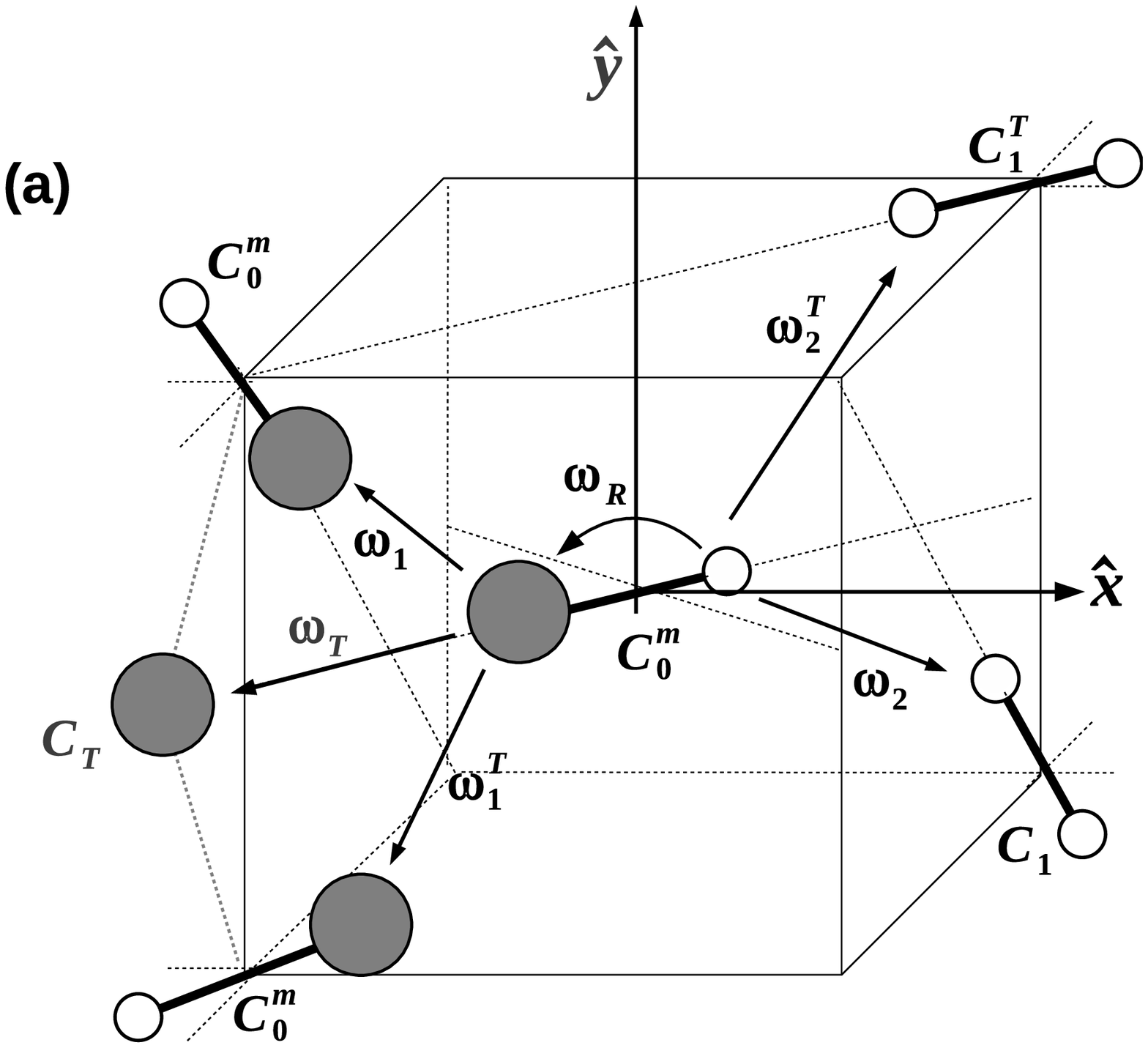}\hspace{-0.4cm}\includegraphics[width=6.cm]{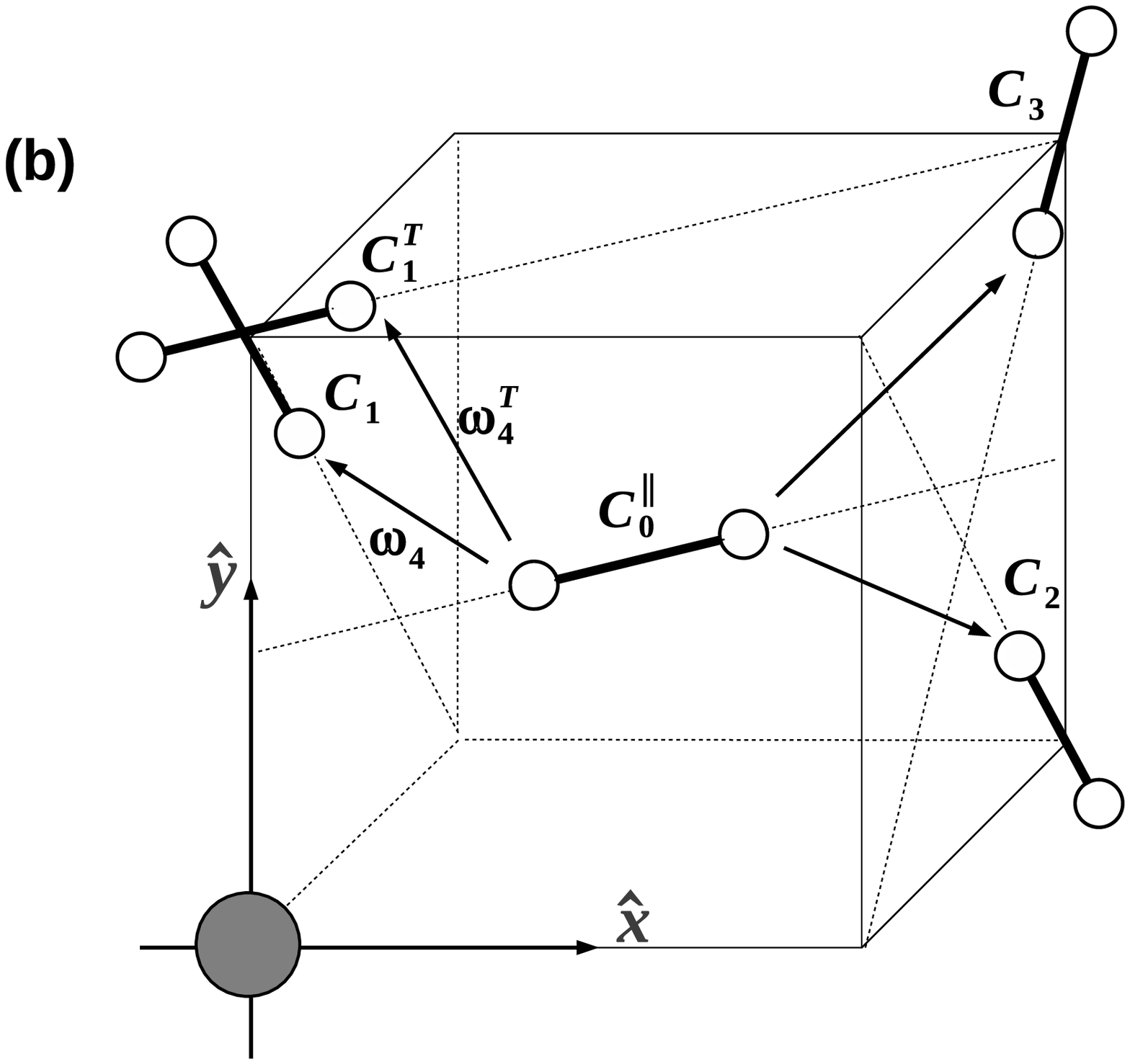}\hspace{-0.4cm}\includegraphics[width=6.cm]{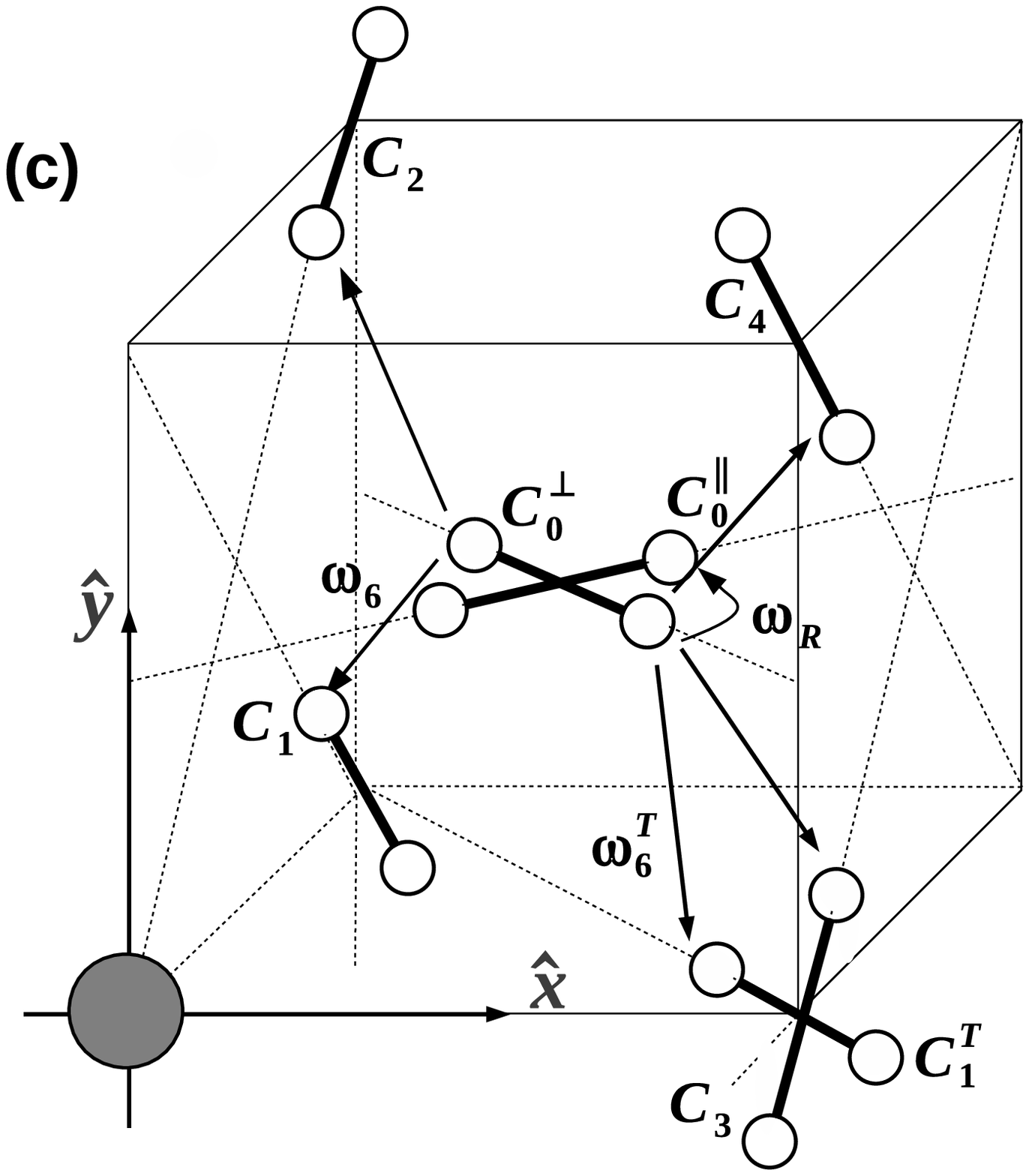}
\caption{Schematic diagrams of the interstitial dumbbells hopping jumps. (a) By translation; (b) by translation-rotation mechanisms. In (c) jumps from the mixed dumb-bell. In the figures, the white and grey circles represent respectively the solvent solute atoms. The migration barriers of the individual jumps are presented in Table \ref{TIjump}.
.}
\label{IntJ}
\end{center}
\end{figure}

Diffusion via interstitial mechanism is characterized by the following jump frequencies: $\omega_i$ for dumb-bell roto-translations and $\omega^T_i$ (with $i=1,\dots,6$) for dumb-bell translations in the context of the second binding model or second shell approximation \cite{BAR07}. For simplicity, we use the same frequency notation as in Ref. \cite{CHO11}. In the phenomenological description, only the above set of frequencies is employed in the full set of $L$-coefficients calculations.

\begin{table}[h]
\begin{center}
\caption{Interstitials migration barriers from CMST calculations. Binding energies $\epsilon_m$ are shown in the first colums. The jumps are depicted in the third column, while the forth and fifth columns describe the migration energies $E^{\leftrightarrow}_m$ for direct and reversed jumps. In the last column, results from Ref. \cite{CHO11} using VASP.}
\label{TIjump}
\begin{tabular}{ccccc|cc} 
\hline 
&&&&&&\\
Figure \ref{IntJ}& $\epsilon^I_m(eV)$ & $j_{\alpha }$ & $E^{\rightarrow}_m(eV)$ & $E^{\leftarrow}_m(eV)$ & $E^{\rightarrow}_m(eV)$ & $E^{\leftarrow}_m(eV)$ \\
&&&&&&\\
\hline 
 &  & 
$\xymatrix{
C_0  \ar@<0.5ex>[r]^{\omega_0}
& C_0 \ar@<0.5ex>[l]^{\omega_0} }
$ &  0.31  & 0.31 & 0.35 & 0.35  \\
&   & 
$\xymatrix{
C_0  \ar@<0.5ex>[r]^{\omega^T_0}
& C_0 \ar@<0.5ex>[l]^{\omega^T_0} }
$ &  0.85  & 0.85 & 0.84 & 0.84 \\ 
\hline
\textbf{(a)} &  &
$\xymatrix{
C^m_0  \ar@<0.5ex>[r]^{\omega_R}
& C^m_0 \ar@<0.5ex>[l]^{\omega^{\prime}_R} }
$ & 0.22 & 0.22 & - & - \\
&  &
$\xymatrix{
C^m_0  \ar@<0.5ex>[r]^{\omega^T_1}
& C^m_0 \ar@<0.5ex>[l]^{\omega^T_1}}
$ & 0.46 & 0.46 & 0.48 & 0.48  \\
& & 
$\xymatrix{
C^m_0  \ar@<0.5ex>[r]^{\omega_1}
& C^m_0 \ar@<0.5ex>[l]^{\omega_1} }$ & 0.22 & 0.22 & 0.25 & 0.25 \\
&  -0.051 &
$\xymatrix{
C^m_0  \ar@<0.5ex>[r]^{\omega_2}
& C_1 \ar@<0.5ex>[l]^{\omega_3} }$ & 0.37 & 0.43 & 0.33 & 0.42 \\
& -0.051 &
$\xymatrix{
C^m_0  \ar@<0.5ex>[r]^{\omega^T_2}
& C^T_1 \ar@<0.5ex>[l]^{\omega^T_3} }$ & 0.56 & 0.67 & 0.58 & 0.68 \\
& 0.044 &
$\xymatrix{
C^m_0  \ar@<0.5ex>[r]^{\omega_T}
& C_T \ar@<0.5ex>[l]^{\omega'_T} }$ & 0.16 & 0.01 & - & - \\
\hline
\textbf{(b)} & -0.124 &
$\xymatrix{
C^{\parallel}_{0}  \ar@<0.5ex>[r]^{\omega^{\prime}_{R}}
& C^{\perp}_{0} \ar@<0.5ex>[l]^{\omega_{R}} }
$ & 0.37 & 0.44 & - & - \\
& -0.033 &
$\xymatrix{
C^{\parallel}_{0}  \ar@<0.5ex>[r]^{\omega _{4}}
& C_{1} \ar@<0.5ex>[l]^{\omega_{5}} }
$ & 0.35 & 0.26 & 0.39 & 0.26 \\
& -0.056&
$\xymatrix{
C^{\parallel}_0 \ar@<0.5ex>[r]^{\omega^T_4}
& C^{T}_1 \ar@<0.5ex>[l]^{\omega^T_5} }
$ & 0.78 & 0.78 & 0.79 & 0.79 \\
& -0.018 &
$\xymatrix{C^{\parallel}_0  \ar@<0.5ex>[r]^{}
& C_2 \ar@<0.5ex>[l]^{}}$ & 0.30 & 0.27 & - & - \\
& -0.002 &
$\xymatrix{
C^{\parallel}_{0}  \ar@<0.5ex>[r]^{}
& C_3 \ar@<0.5ex>[l]^{}}
$ & 0.27 & 0.22 & - & - \\
\hline
\textbf{(c)} & -0.051 &
$\xymatrix{
C^{\perp}_0  \ar@<0.5ex>[r]^{\omega_R}
& C^{\parallel}_0 \ar@<0.5ex>[l]^{\omega^{\prime}_R} }
$ & 0.44 & 0.37 & - & - \\
& -0.033 &
$\xymatrix{
C^{\perp}_0  \ar@<0.5ex>[r]^{\omega_6}
& C_1 \ar@<0.5ex>[l]^{\omega_7}}
$ & 0.36 & 0.42 & 0.36 & 0.37  \\
& -0.033 &
$\xymatrix{
C^{\perp}_0  \ar@<0.5ex>[r]^{\omega^T_6}
& C^T_1 \ar@<0.5ex>[l]^{\omega^T_7} }$ & 0.32 & 0.36 & 0.32 & 0.35 \\
& -0.102 &
$\xymatrix{
C^{\perp}_0  \ar@<0.5ex>[r]^{}
& C_2 \ar@<0.5ex>[l]^{}}$ & 0.31 & 0.28 & - & - \\
& -0.056 &
$\xymatrix{
C^{\perp}_0  \ar@<0.5ex>[r]^{}
& C_3 \ar@<0.5ex>[l]^{} }$ & 0.32 & 0.25 & - & - \\
& -0.018 &
$\xymatrix{
C^{\perp}_0  \ar@<0.5ex>[r]^{}
& C_4 \ar@<0.5ex>[l]^{} }$ & 0.32 & 0.22 & - & - \\
\hline
\end{tabular}
\end{center}
\end{table}

The full set of frequencies corresponding to pure translation and roto-translation of the dumb-bell at nearest neighbor sites from the solute $Cr$, are calculated from values of $E^{\rightarrow}_m$ in Table \ref{TIjump}. As for vacancies, if we assume that the jumps are thermally activated then, $\omega_{i}$ calculated from expressions (\ref{nu0T}) and (\ref{nuT}).

Binding energies, using CMST calculations, are displayed in Table \ref{TIjump}. In table are also shown the different type of $Cr$-interstitial complex $C_k=S+I_k$ with its binding energies $\epsilon^I_m$, together with the possibles configurations and jumps that involve the corresponding $C_k=S+I_k$ complex with the corresponding jump frequencies. Interstitial migration barriers are shown for the direct as well as for the reverse jumps. Results from CMST shown a weak attractive energy interaction, $\epsilon^I_m$, between the interstitial and solute in all pairs configurations with the exception of the $C^{\parallel}_0$ minimum configuration, which have a strong binding energy with $Cr$, although the minimum absolute relaxed configurations corresponds to the $C^{\perp}_0$ one. Concerning to the migration barriers calculated with the Monomer in CMST in Table \ref{T5}, we see that, the interstitial migration barriers $E^{\rightarrow}_m$ and $E^{\leftarrow}_m$ are quite different from that in perfect $Fe$ lattice, $E^I_m=0.31,85eV$, also confirmed (not shown) by DFT calculations. We remark that, in Table \ref{TIjump} our CMST present calculations and DFT from Ref. \cite{CHO11} using VASP, gives practically same results, and consequently same frequencies. Then, we only present our result obtained from CMST calculations for a temperature dependent attempt frequency $\nu_0(T)$ from (\ref{nuT}).  
\clearpage
\begin{figure}[h]
\begin{center}
\includegraphics[angle=-90,width=8.50cm]{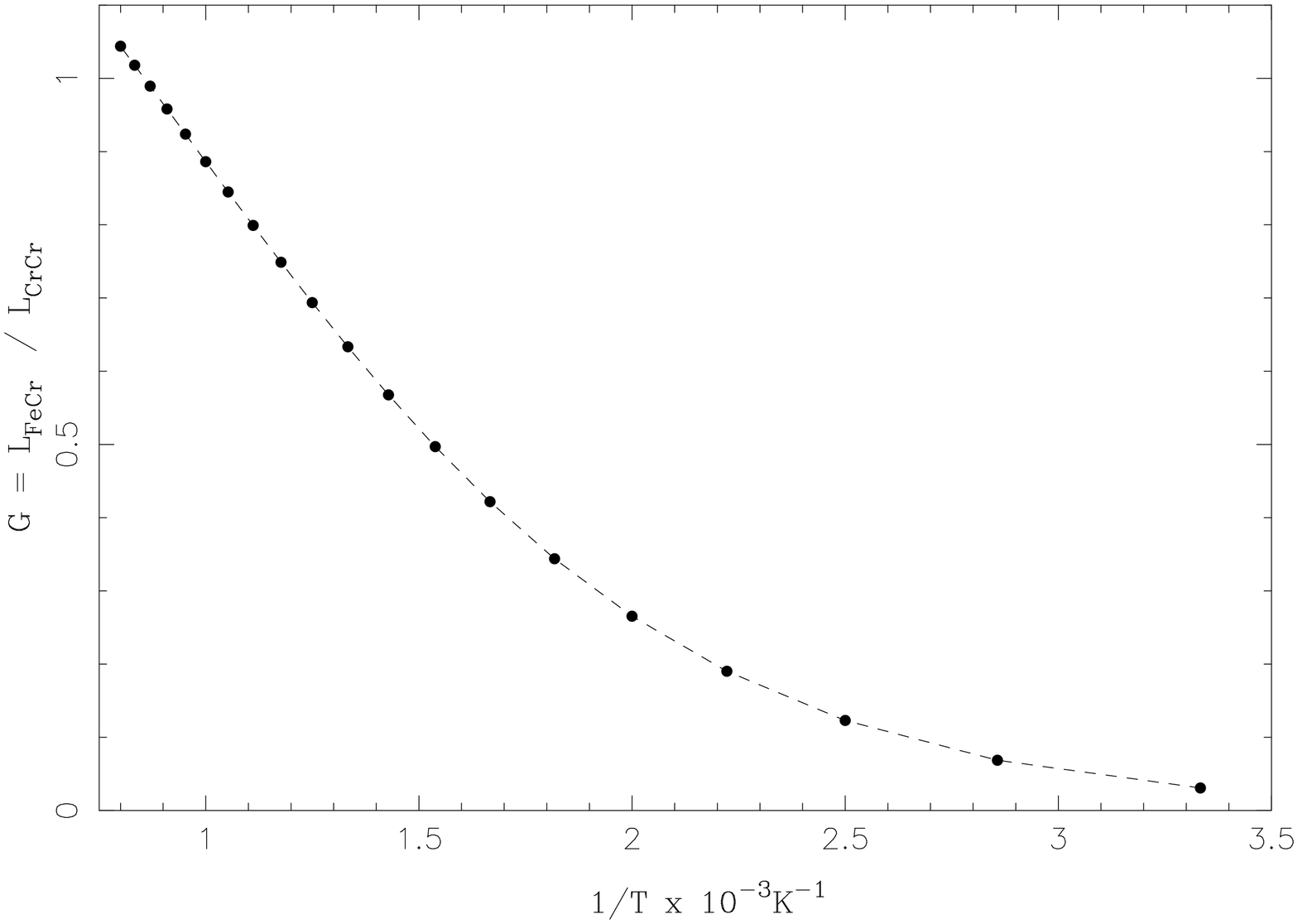}
\vspace{1.3cm}
\caption{Ratio of the interstitial-Onsager coefficients of $Cr$ in $Fe$ calculated from expressions $L_{FeCr}$ and $L_{CrCr}$ in Ref. \cite{BAR07} \textit{vs} $1/T$..}
\label{FIG17-I}
\end{center}
\end{figure}

\begin{figure}[h]
\begin{center}
\includegraphics[angle=-90,width=8.50cm]{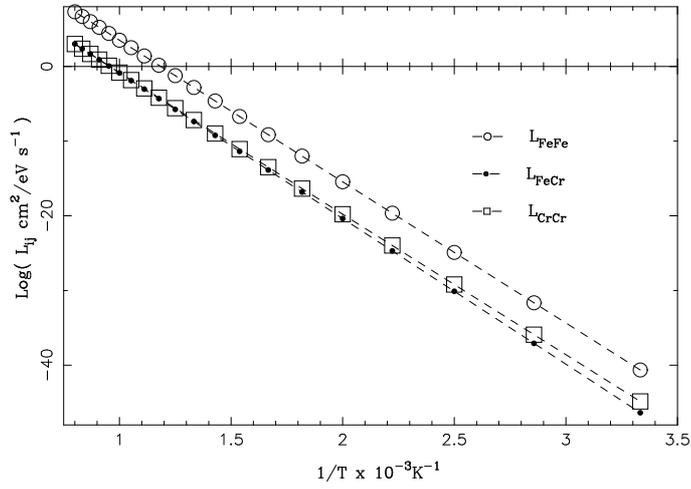}
\vspace{1.3cm}
\caption{Interstitial-Onsager coefficients \textit{vs} $1/T$ for the $FeCr$ system. Squares denote $L_{CrCr}$, empty circles denote $L_{FeFe}$ while $L_{FeCr}$ is described with filled circles.}
\label{FIG18-I}
\end{center}
\end{figure}

\begin{figure}[h]
\begin{center}
\includegraphics[angle=-90,width=8.50cm]{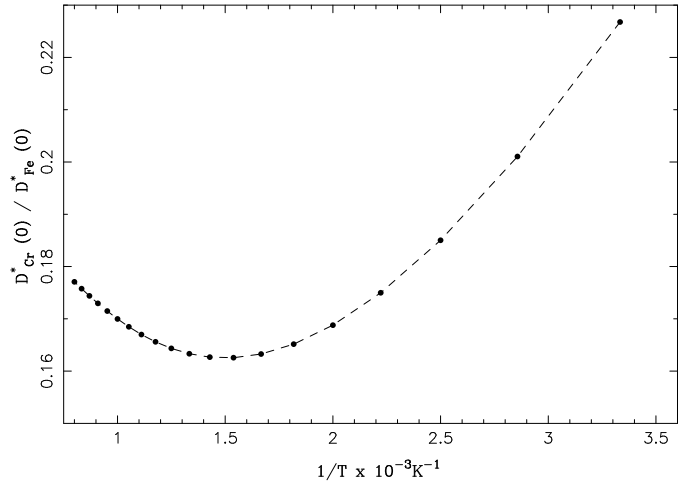}
\vspace{1.3cm}
\caption{Ratio of the tracer diffusion coefficient $D^{\star}_{Cr}/D^{\star}_{Fe}$ in $FeCr$ \textit{vs} $1/T$.}
\label{FIG19-I}
\end{center}
\end{figure}
\begin{figure}[h]
\begin{center}
\hspace{-0.5cm}\includegraphics[angle=-90,width=8.0cm]{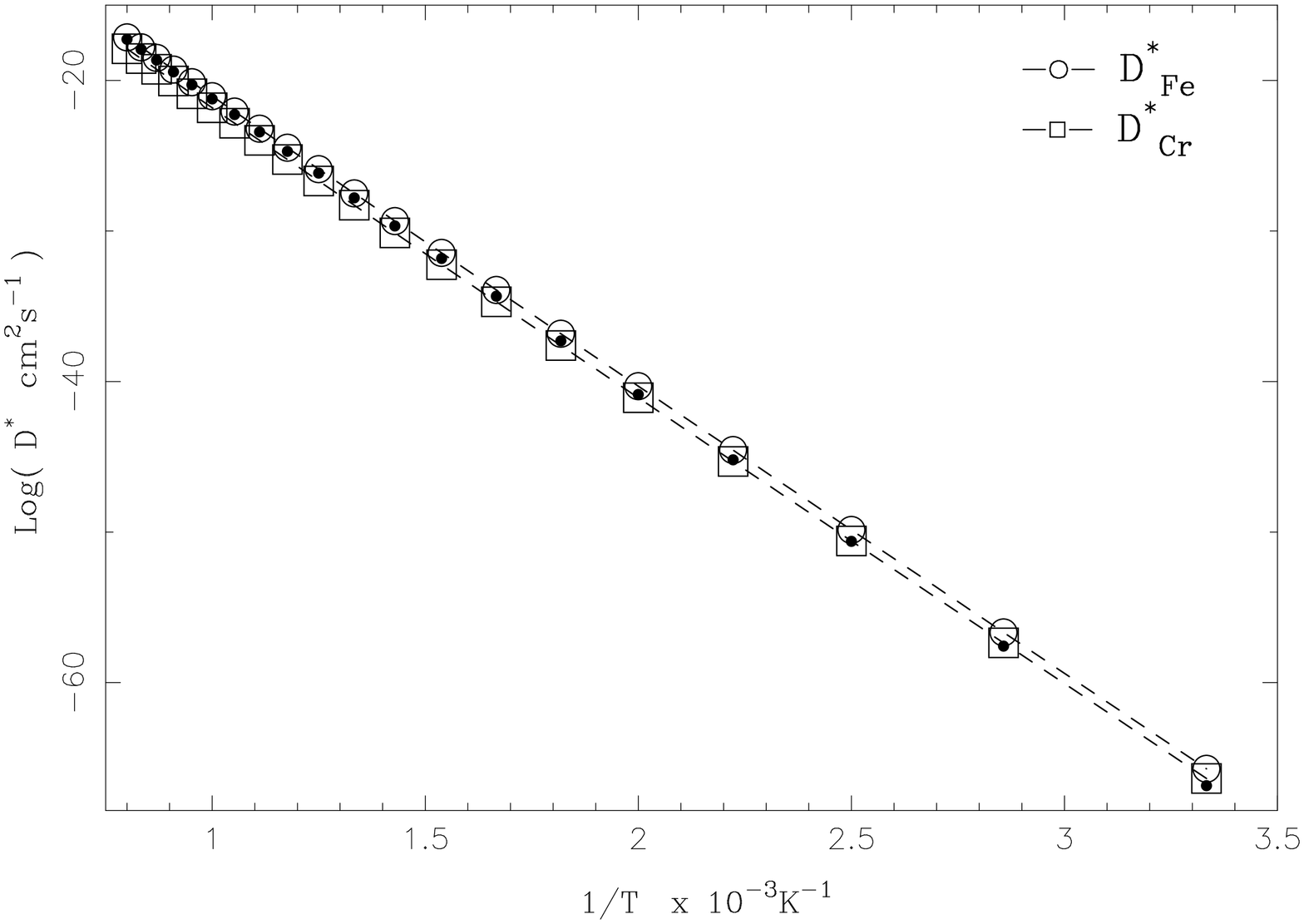}\hspace{0.55cm}\includegraphics[angle=-90,width=8.0cm]{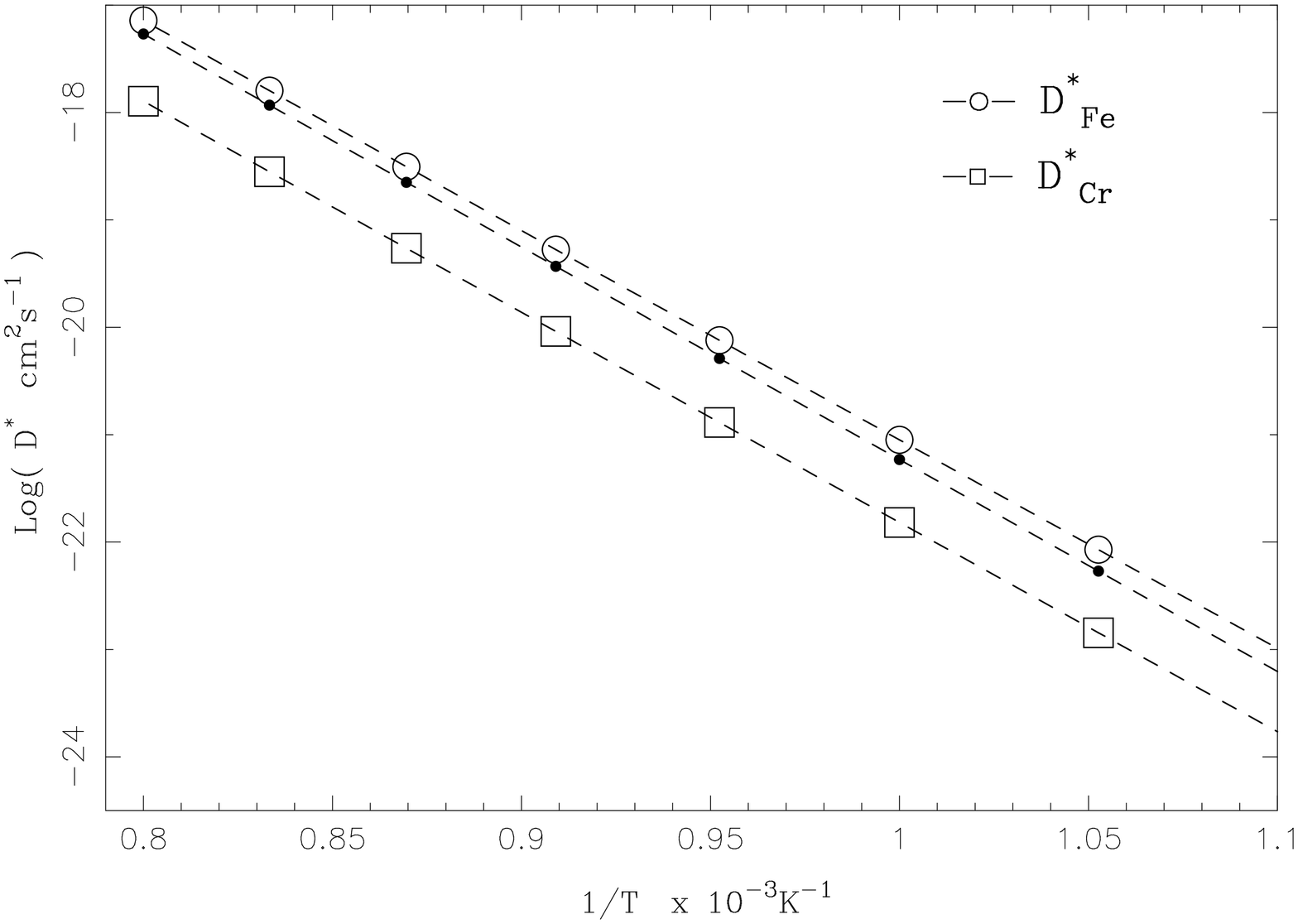}
\vspace{1.3cm}
\caption{Tracer solute diffusion coefficients, $D^{\star}_{Cr}$, in open squares, while self diffusion coefficient $Fe$, $D^{\star}_{Fe}$, in open circles. Black filled circles correspond to the tracer self diffusion coefficient $D^{\star}_{Fe}$ calculated from the intrinsic diffusion coefficient \cite{ALL03}.}
\label{FIG20-I}
\end{center}
\end{figure}

From the calculated $L_{ij}$ transport coefficients using expressions in Ref. \cite{BAR07}, the tracer solute diffusion coefficient, via the interstitial mechanism for $Cr$, can be calculated as a function of the temperature using an expression that connect the jump rates in Figure \ref{IntJ} and macroscopic transport coefficient as,
\begin{equation}
D^{\star}_{Cr}=\frac{k_BT}{n_{Cr}}L_{CrCr},
\end{equation}
where $n_{Cr}$ is the number of $Cr$ atoms per unit volume. The results are plotted in Fig. \ref{FIG20-I}. The tracer self-diffusion coefficient $D^{\star}_{Fe}$, is calculated
by evaluating the transport coefficient $L_{CrCr}$ with every migration frequency set equal to the value for pure $Fe$. In other words, every $\omega_i$ and $\omega^T_i$ are replaced by $\omega_0$ and $\omega^T_0$, as in Ref. \cite{CHO11}. 
Black filled circles in Figure \ref{FIG20-I} corresponds to the tracer self diffusion coefficient, $D^{\star}_{Fe}$, calculated from the intrinsic diffusion coefficient in terms of $L_{FeFe}$ and $L_{FeCr}$ from non-equilibrium thermodynamics, and the flux equations assuming that $Cr$ can only move move by the action of interstitials \cite{ALL03}. 

\clearpage
\section{Concluding remarks}
\vspace{0.5cm}
In summary, in this work the diffusion properties in BCC $FeCr$ diluted alloys have been studied.
The diffusion processes here considered can be mediated by vacancies as well as by interstitials.

The flux equations permits to relates the diffusion coefficients with the Onsager tensor. Kinetic theory allows to write this Onsager coefficients in terms of jump frequencies. In this way we could write expressions for the diffusion coefficients only in terms of microscopic magnitudes, i.e. the jump frequencies.

In this context, for the case of a diffusion process mediated by vacancies, in this work we use for the first time in a real system, the approach  developed by Okamura and Allnatt \cite{OKA83}.
This procedure, is known as the  six frequencies model and involves a second nearest neighbor binding approach. On the other hand for the case of the diffusion mediated by instestitials, the recent procedure developed by Barbe and Nastar  \cite{BAR07} is here employed.

The needed frequencies involved in the Onsager coefficients for both vacancies and interstitial mechanisms have been calculated thanks to the economic technique namely the Monomer method coupled to CMST and DFT calculations, this last one using SIESTA. The results employing either CMST or DFT  calculations are similar, which confirms the advantage of the lower cost CMST methods that were already reported in \cite{RAM13,RAM14}.

 The vacancy tracer diffusion coefficient for $FeCr$ were also compared with available experimental data obtaining a good agreement with the here described theory. However in comparison with results obtained within the first coordinated shell approach \cite{CHO11} the diffusion coefficients here described up to second coordinate neighbor are several orders of magnitudes higher, showing  the need for the description of the diffusion process beyond the first  shell approximation. Also the present CMST calculations, confirmed by DFT results, reveal that $Cr$ in $Fe$ at diluted concentrations migrates as free species which implies that a vacancy drag mechanism is unlikely to occur in BCC $FeCr$ within the present approach.

For the case of the diffusion coefficients mediated by interstitial mechanism, although the results obtained here are in accordance with those in \cite{CHO11} they are four orders of magnitudes lower than the vacancy diffusion coefficients and than experimental data. Hence we can conclude that for the $FeCr$ diluted alloys the diffusion process is mainly due to a vacancy mechanism.   

Recently, Garnier \textit{et al.} \cite{GAR13}, have obtained analytic expressions for diffusion coefficients in binary alloys with BCC structure. The authors have investigated the drag of solute atoms by vacancies at low temperatures, using a self-consistent mean field method. The method takes into account interactions between the solute atom and a vacancy up to the third nearest neighbor sites. Analytic results have identified the mechanism involved in the solute drag by vacancies. Then, it would be interesting to extend the calculations for a twelve-frequency model as in Ref. \cite{GAR13} in order to obtain convergent results independently of the approximation employed.

\section*{Acknowledgements}
I am grateful to Dr. A.M.F. Rivas for comments on the manuscript and to Mart\'in Urtubey for Figure \ref{6f-V}. This work was partially financed by CONICET PIP-00965/2010 and the CNEA/CAC - Gerencia Materiales.

\bibliographystyle{elsarticle-num}

\end{document}